\documentclass[aps,prd,preprintnumbers,showpacs,superscriptaddress,nofootinbib,amsmath,amssymb,floats,floatfix,showkeys,notitlepage,longbibliography]{revtex4-1}
\usepackage{comment}
\usepackage{graphicx}
\usepackage{subfigure}
\usepackage{palatino}
\usepackage[commandnameprefix=always]{changes}
\usepackage{hyperref}
\hypersetup{colorlinks=true,linkcolor=orange,urlcolor=blue,citecolor=red}
\usepackage[toc,page]{appendix}
\usepackage[normalem]{ulem}
\usepackage{orcidlink}
\usepackage{lipsum}
\usepackage{graphicx}
\usepackage{subfigure}
\usepackage{palatino}
\usepackage{float}
\usepackage{sans}
\usepackage{adjustbox}
\usepackage{latexsym}
\usepackage{amsmath}
\usepackage{amssymb}
\usepackage{amsfonts}
\usepackage{dcolumn}
\usepackage{bm}
\usepackage{tikz}
\usepackage{bigints}
\usepackage{array,tabularx,multirow,booktabs}
\usepackage[tracking=true]{microtype}
\SetTracking{}{500}
\SetTracking{encoding={*}, shape=sc}{40}
\allowdisplaybreaks
\usepackage{adjustbox}
\usepackage{latexsym}
\usepackage{amsmath}
\usepackage{amssymb}
\usepackage{amsfonts}
\usepackage{dcolumn}
\usepackage{bm}
\usepackage{tikz}
\usepackage{mathrsfs}
\usepackage{bigints}
\usepackage{array,tabularx,multirow,booktabs}
\usepackage[tracking=true]{microtype}
\usepackage{color}
\allowdisplaybreaks

\begin{document}
\title{{\bf \LARGE Thermodynamic Curvature and Topological Insights of Hayward Black Holes with String Fluids \\\;\\ \hrule}
}
\vspace{1cm}
\author{ \textbf{Ankit Anand\orcidlink{0000-0002-8832-3212}$^\ddag$, Saeed Noori Gashti\orcidlink{0000-0001-7844-2640}$^\dag$ and Aditya Singh\orcidlink{0000-0002-2719-5608}$^\star$}}
\email{ Anand@iitk.ac.in, saeed.noorigashti@stu.umz.ac.ir; sn.gashti@du.ac.ir; saeed.noorigashti70@gmail.com, 24pr0148@iitism.ac.in}
\vspace{1.5cm}
\affiliation{\normalsize\it $^\ddag$ Department of Physics, Indian Institute of Technology Kanpur, Kanpur 208016, India. \\
$^\dag$ School of Physics, Damghan University, P. O. Box 3671641167, Damghan, Iran. \\
$^\star$ Department of Physics, Indian Institute of Technology (Indian School of Mines), Dhanbad,
Jharkhand-826004, India
}

\begin{abstract}
\vspace{1cm}
\begin{center}
    \textbf{Abstract}
\end{center}
In this paper, we study the influence of string fluids on the extended thermodynamic structure and microscopic interactions of Hayward black holes by employing thermodynamic geometry as an empirical tool. Using the novel equation of state obtained for regular black holes surrounded by string fluids, we analyze the extended phase space with enthalpy as the central thermodynamic potential. By examining the behavior of the normalized Ruppeiner curvature scalar $R_N$ in the temperature-volume $(T,V)$ plane, we analyzed the influence of the string fluid parameters on the microstructure of the black hole. Our analysis reveals that the presence of string fluids significantly modifies the dominant microscopic interactions, transitioning from attractive to repulsive regimes depending on the charge and volume of the black hole. We see that the thermodynamic curvature effectively detects critical points and phase transitions, reflecting the nature of repulsive or attractive interactions among black hole microstructures. We further investigate thermodynamic topology to provide a novel classification scheme for stability and phase behavior, delineating local stable and unstable regions in parameter space. We investigate the thermodynamic topology of Hayward-AdS black holes surrounded by string fluids, showing that the number and type of topological charges depend on the parameters $\epsilon$ and $b$, revealing phase transitions and stability characteristics encoded in the global topological charge $W$. This integrated study of the thermodynamic geometry and topology structure enhances the understanding of Hayward black holes surrounded by string fluids, showing overall thermodynamic stability and configuration with significant implications for holographic duality and potential astrophysical observations.

\vspace{1cm}
\end{abstract}

\maketitle
\tableofcontents


\section{Introduction}\label{sec-intro}

Black hole thermodynamics has emerged as a pivotal framework at the intersection of gravitation, quantum theory, and statistical mechanics. The field gained prominence following the discovery that the variation of a black hole’s mass exhibits a striking analogy with the first law of thermodynamics, thereby attributing thermodynamic properties to these enigmatic objects~\cite{Bardeen:1973gs}. In the semiclassical regime \cite{Bekenstein:1973ur, Hawking:1975vcx}, black holes are characterized by a temperature proportional to their surface gravity and an entropy proportional to the area of their event horizon, with Newton’s gravitational constant serving as the proportionality factor. These foundational insights have laid the groundwork for extensive investigations into the thermodynamic behavior of gravitational systems. Asymptotically Anti-de Sitter (AdS) black holes, in particular, have received heightened attention due to their role in the AdS/CFT correspondence, where gravitational dynamics in the bulk AdS spacetime are holographically dual to thermal states in a conformal field theory (CFT) defined on the boundary \cite{Maldacena:1997re}. A well-known manifestation of this duality is the Hawking–Page phase transition, which describes a thermodynamic shift between a large black hole and a thermal radiation background in AdS space. This transition corresponds to a confinement/deconfinement phase change in the boundary CFT \cite{Hawking:1982dh}. Through this holographic perspective, black holes serve as powerful tools for probing complex quantum phenomena, including intricate phase structures such as triple points analogous to those predicted in the quantum chromodynamics (QCD) phase diagram \cite{Cong:2021jgb, Cui:2021qpu}.

The presence of singular solutions is often interpreted as signaling the breakdown of classical General Relativity, underscoring the necessity for a more fundamental quantum theory of gravity to resolve such pathologies \cite{Bambi:2013ufa}. Motivated by these considerations, the exploration of singularity-free, or regular, black hole metrics has garnered considerable attention the first phenomenologically motivated regular black hole solution \cite{Barca:2023shv}. Among these, the Hayward metric stands out as a prototypical example of a static, spherically symmetric regular black hole that asymptotically approaches de Sitter spacetime near the origin ($ r\to 0$) and recovers asymptotic flatness at spatial infinity ($ r\to \infty$) \cite{Dymnikova:1992ux}. This solution provides a physically compelling model of black hole formation from an initial vacuum state endowed with finite density and pressure distributions that vanish asymptotically. Notably, the Hayward geometry admits an interpretation as an exact solution of Einstein’s equations coupled to nonlinear electrodynamics. The gauge-invariant model of a string cloud, inspired by string-theoretic considerations wherein fundamental constituents are one-dimensional extended objects rather than point particles, has enriched the landscape of gravitational solutions \cite{PhysRevD.20.1294}. This framework was further extended to encompass a fluid of strings characterized by an effective pressure component \cite{Letelier}. A static black hole immersed in such a string fluid revealed corrections to the Newtonian gravitational force scaling as $1/r$, offering a potential phenomenological avenue for explaining galactic rotation curves \cite{Soleng1995, Soleng1994}.

Thermodynamic curvature, particularly in the form of Ruppeiner geometry~\cite{Ruppeiner1995, Ruppeiner2010}, and thermodynamic topology are deeply interconnected tools used to analyze phase transitions and microstructures of thermodynamic systems, especially black holes. Ruppeiner geometry introduces a Riemannian metric on the thermodynamic state space by taking the negative Hessian of the entropy with respect to extensive variables, and its scalar curvature $R$ provides insights into the nature of microscopic interactions. Thermodynamic topology, on the other hand, uses global invariants such as the Euler characteristic and Betti numbers to classify different thermodynamic phases \cite{Zhang2020}. The two approaches connect through the Gauss–Bonnet theorem, where integrating the thermodynamic curvature over a manifold can yield topological invariants, thus linking local geometric features to global topological structures \cite{Banerjee2022}. In black hole thermodynamics, for instance, divergences in $R$ at phase transitions—such as the Hawking–Page or van der Waals-like transitions—can correspond to changes in the topology of the thermodynamic state space, reflecting a shift in the number or nature of critical points \cite{Wei2019, Yerra2022}. Therefore, thermodynamic curvature provides a local geometric probe of interactions and stability, while thermodynamic topology offers a global classification framework, and together they form a powerful unified picture for studying the phase structure and underlying microphysics of complex thermodynamic systems.


\section{Review of Hayward Black Holes Surrounded by String Fluids}

We briefly discuss the Hayward black holes surrounded by string fluids. We begin our analysis by considering the Hayward black hole spacetime \cite{hayward2006formation, nascimento2024some}, which arises as a solution to Einstein’s field equations coupled to a nonlinear electromagnetic field. The physical source of the gravitational field is a nonlinear electrodynamics (NED) configuration \cite{molina2021thermodynamics, bronnikov2001regular, fan2016construction}. The action for the system, minimally coupled to a nonlinear electromagnetic field \cite{bronnikov2001regular, fan2016construction, molina2021thermodynamics} as 
\begin{equation}
\mathcal{S} = \frac{1}{16\pi}\int d^4x\sqrt{-g} \left(R+\mathcal{L}(\mathcal{F}) \right) \;\;\; \text{where} \;\;\mathcal{L}(\mathcal{F}) = \frac{6 \left(2\mathrm{h}^2\mathcal{F} \right)^{3/2}}{\kappa^2 \mathrm{h}^2\left[1+\left(2\mathrm{h}^2\mathcal{F}\right)^{3/4}\right]^2}
\label{eq_action}
\end{equation}
where $g$ is the determinant of the metric tensor $g_{\mu\nu}$, $R$ denotes the Ricci scalar, $\mathrm{h}$ is the Hayward parameter, typically associated with a fundamental length scale, and $\kappa^2 = 8\pi$. The Lagrangian is a nonlinear function of the electromagnetic invariant $\mathcal{F}=F^{\mu\nu}F_{\mu\nu}$. Varying the action (\ref{eq_action}) with respect to the metric tensor yields the modified Einstein field equations \cite{bronnikov2001regular}
\begin{equation}
R_{\mu\nu} - \frac{1}{2}R\,g_{\mu\nu} = 2 \left(\frac{\partial\mathcal{L}(\mathcal{F})}{\partial F}F_{\mu\sigma}F^{\sigma}{_\nu}-\frac{1}{4}g_{\mu\nu}\mathcal{L}(\mathcal{F})\right) \ .
\label{equação de einstein mista}
\end{equation}

For a purely magnetic configuration in a spherically symmetric spacetime, the only non-vanishing component of $F_{\mu\nu}$ is computed in \cite{bronnikov2001regular} as $F_{23} = q_m \sin{\theta}$ and leading to the field invariant
\begin{equation}
F = \frac{2q_m^2}{r^4} \ .
\label{F}
\end{equation}
Hayward suggested that the parameter $l$ is of the order of the Planck length. This parameter can be related to the magnetic charge $q_m$ through the relation \cite{molina2021thermodynamics, bronnikov2001regular, fan2016construction, bronnikov2017comment, toshmatov2018comment}
\begin{equation}
q_m = \frac{\sqrt[3]{4m^2 l}}{2} \ ,
\label{magnetic charge}
\end{equation}
Using this configuration, the components of the stress-energy tensor associated with the Hayward solution take the form \cite{hayward2006formation}
\begin{equation}
T_t^{t} = T_r^{r} = \frac{12 \mathrm{h}^2 m^2}{(r^3 + 2 \mathrm{h}^2 m)^2}\;\;\;;\;\;\; T_\theta^{\theta} = T_\phi^{\phi} = -\frac{24 (r^3 - \mathrm{h}^2 m) \mathrm{h}^2 m^2}{(r^3 + 2 \mathrm{h}^2 m)^3} \ ,
\label{eq:1.6}
\end{equation}
where $l$ and $m$ are assumed to be positive constants. For incorporating two additional physical sources: a cosmological constant and a fluid of strings, the EOM \eqref{equação de einstein mista} by adding the term $-\Lambda g_{\mu\nu}$ to the left-hand side and introduce an additional energy-momentum contribution $T_{\mu\nu}^{\text{FS}}$ on the right-hand side, corresponding to the string fluid.

For the inclusion of the fluid of strings, one starts with the trajectory of a point particle with four-velocity $u^\mu = dx^\mu/d\lambda$, where $\lambda$ is an arbitrary parameter, represented by the worldline $x^\mu = x^\mu(\lambda)$. In contrast, for an infinitesimally thin relativistic string, the trajectory traces out a two-dimensional worldsheet $\Sigma$, which can be parametrized as \cite{Letelier}
\begin{equation}
x^\mu = x^\mu(\lambda^a), \quad a = 0, 1,
\label{eq:1.10}
\end{equation}
where $\lambda^0$ and $\lambda^1$ denote, respectively, timelike and spacelike parameters on the worldsheet. The bivector $\Sigma^{\mu\nu}$ is defined as
\begin{equation}
\Sigma^{\mu\nu} = \epsilon^{ab} \frac{\partial x^\mu}{\partial \lambda^a} \frac{\partial x^\nu}{\partial \lambda^b},
\label{eq:1.11}
\end{equation}
where $\epsilon^{ab}$ is the two-dimensional Levi-Civita symbol. The worldsheet is equipped with an induced metric $\gamma_{ab}$, given by
\begin{equation}
\gamma_{ab} = g_{\mu\nu} \frac{\partial x^\mu}{\partial \lambda^a} \frac{\partial x^\nu}{\partial \lambda^b},
\label{eq:1.12}
\end{equation}
with determinant $\gamma = \det(\gamma_{ab})$. The stress-energy tensor for a dust-like distribution of particles is expressed as $T^{\mu\nu} = \rho u^\mu u^\nu$, where $\rho$ is the rest-frame energy density and $u^\mu$ is the normalized four-velocity. Analogously, for a cloud of strings, the corresponding stress-energy tensor \cite{Letelier} is given by 
\begin{equation}
T^{\mu\nu} = \rho \frac{\Sigma^{\mu\beta} \Sigma_\beta^{\nu}}{\sqrt{-\gamma}}\;\;\;\;;\;\;\text{where}\;\; \gamma = \frac{1}{2} \Sigma^{\mu\nu} \Sigma_{\mu\nu}\ .
\label{eq:1.13}
\end{equation}
In the case of a perfect fluid with pressure $p$, the stress-energy tensor assumes the standard form $T^{\mu\nu} = (\rho + p) u^\mu u^\nu - p g^{\mu\nu}$. For a perfect fluid composed of strings, including pressure effects, the energy-momentum tensor\cite{Letelier} is generalized to 
\begin{equation}
T^{\mu\nu} = \left(p + \sqrt{-\gamma}\rho\right) \frac{\Sigma^{\mu\beta} \Sigma_\beta^{\nu}}{-\gamma} + p g^{\mu\nu} \ .
\label{EMT for String}
\end{equation}
Utilizing the energy-momentum tensor in Eq.~\eqref {EMT for String}, the components of this tensor \cite{soleng1995dark} satisfy the following relations
\begin{equation}
T_t^{t} = T_r^{r}\;\;\;\;;\;\;\;\;T_t^{t} = -\beta \,T_\theta^{\theta} = -\beta T_\phi^{\phi} \ ,
\label{eq:1.15}
\end{equation}
where $\beta$ is a dimensionless constant parameterizing the anisotropy of the fluid. These relations characterize an anisotropic fluid configuration with spherical symmetry \cite{dymnikova1992vacuum, soleng1994correction}, and have also been interpreted in the context of vacuum polarization effects in curved spacetimes. Such a form of the stress-energy tensor has been explored in various physical scenarios \cite{salgado2003simple, giambo2002anisotropic, dymnikova2002cosmological}. By adopting the energy-momentum tensor \cite{toledo2020black} for the string fluid is
\begin{equation}
T_t^{t} = T_r^{r} = -\frac{\epsilon}{r^2} \left(\frac{b}{r}\right)^{2/\beta}\;\;\;\;\;;\;\;\;\;\; T_\theta^{\theta} = T_\phi^{\phi} = \frac{\epsilon}{\beta r^2} \left(\frac{b}{r}\right)^{2/\beta} \ ,
\label{eq:1.17.1}
\end{equation}
where $b > 0$ is an integration constant and $\epsilon = \pm 1$ determines the sign of the energy density of the string fluid. 

The form of the line element describing a Hayward-AdS black hole surrounded by a string fluid, as discussed in \cite{Nascimento:2024maj}, is 
\begin{equation}
\begin{aligned}
f(r) = 1 - \frac{2Mr^2}{r^3 +q^3} - \frac{\Lambda r^2}{3} +
\left\{
\begin{array}{ll}
\displaystyle \frac{\epsilon b[1 + 2 \ln(r)]}{2r}, & \beta = 2, \\
\displaystyle \epsilon \beta (\beta - 2)^{-1} \left( \frac{b}{r} \right)^{2/\beta}, & \beta \neq 2.
\end{array}
\right.
\label{eq:f_r}
\end{aligned}
\end{equation}
Finally, it is important to emphasize that this class of solutions can be interpreted as arising from an effective energy-momentum tensor composed of two distinct fluid contributions: one associated with a nonlinear electrodynamic source and another with a string fluid, together with a cosmological constant. By computing the Kretschmann scalar \cite{Nascimento:2024maj}, it is confirmed that for the range $-1 < \beta < 0$, the metric remains regular at the origin. The metric function takes the form
\begin{align}\label{Metric f_r}
f(r) = 1 - \frac{2 M r^2}{r^3 + q^3} - \frac{\Lambda r^2}{3} + \frac{\epsilon\beta}{\beta - 2} \left( \frac{b}{r} \right)^{2/\beta} \ , 
\end{align}

\noindent where $\epsilon$, $\beta$, and $b$ are the model parameters.

\begin{figure}[h!]
    \begin{center}
        \includegraphics[scale=.60]{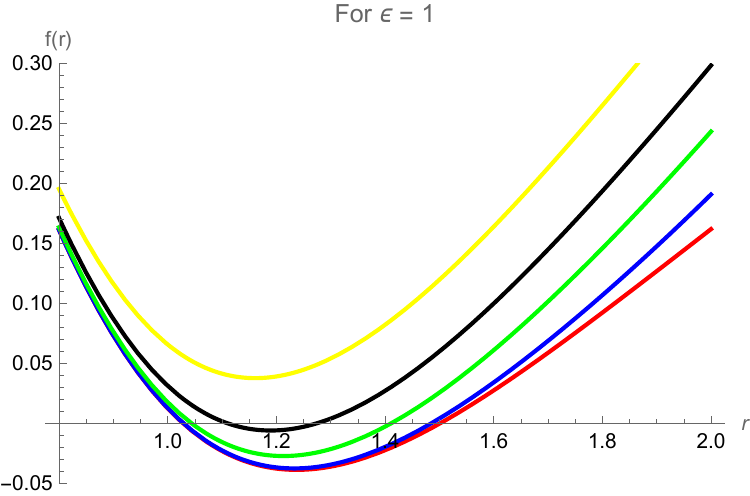} 
        \includegraphics[scale=.60]{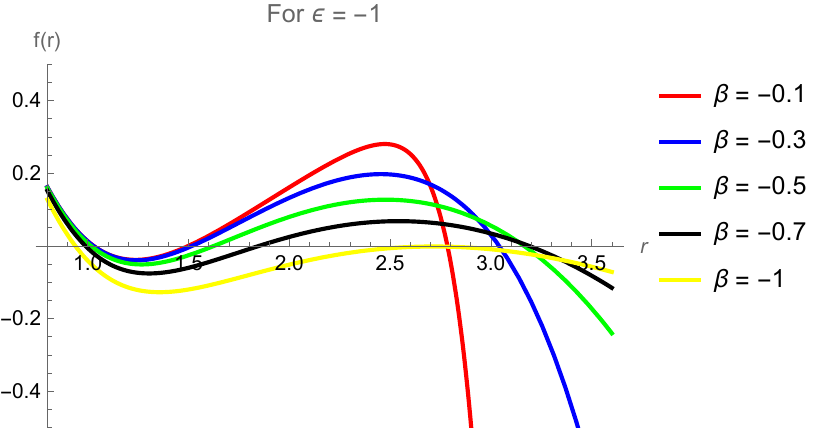} 
    \end{center}
    \caption{The behavior of the Metric function.}
    \label{Metric Plot}
\end{figure}

The identification of the horizon radii requires solving the equation $f(r) = 0$. Due to the complexity of the metric function, closed-form analytical solutions for the horizon radius are not attainable. As a result, we adopt a numerical approach to analyze the horizon structure. Specifically, we examine the behavior of $f(r)$ by plotting it against the radial coordinate $r$, as shown in Figure \ref{Metric Plot}.

To analyze the potential presence of singularities in the spacetime, we examine the behavior of coordinate-invariant quantities constructed from the curvature tensor—most notably, the Kretschmann scalar, defined as $\mathcal{K} = R_{\alpha\beta\mu\nu} R^{\alpha\beta\mu\nu}$. The Krestchmann scalar using Eq.~\eqref{Metric f_r} is
\begin{eqnarray}
   && \mathcal{K} = \frac{8 m^2 \left(r^3-4 l^2 m\right)^2}{\left(2 l^2 m+r^3\right)^4}+\frac{1024 l^6 m^5 \left(l^2 m-r^3\right)}{\left(2 l^2 m+r^3\right)^6}+\frac{8 m^2 r^6 \left(444 l^4 m^2-44 l^2 m r^3+5 r^6\right)}{\left(2 l^2 m+r^3\right)^6}+\frac{16 m \epsilon  \left(\frac{b}{r}\right)^{2/\beta } \left(4 l^2 m-r^3\right)}{(\beta -2) \left(2 l^2 m r+r^4\right)^2} \nonumber \\
    &&+\frac{8 \epsilon ^2 \left(\frac{b}{r}\right)^{4/\beta }}{(\beta -2)^2 r^4}-\frac{32 m \epsilon  \left(\frac{b}{r}\right)^{2/\beta } \left(4 l^4 m^2-14 l^2 m r^3+r^6\right)}{(\beta -2) \beta  r^2 \left(2 l^2 m+r^3\right)^3}-\frac{16 m r \epsilon  \left(\frac{b}{r}\right)^{2/\beta } \left(4 (\beta -4) l^2 m+(\beta +2) r^3\right)}{(\beta -2) \left(2 l^2 m+r^3\right)^3}-\frac{8 (\beta -1) \Lambda  \epsilon  \left(\frac{b}{r}\right)^{2/\beta }}{3 \beta  r^2} \nonumber \\
    && -\frac{64 (\beta -1) l^4 m^3 \epsilon  \left(\frac{b}{r}\right)^{2/\beta }}{(\beta -2) r^2 \left(2 l^2 m+r^3\right)^3}+\frac{4 \left(\beta ^4+3 \beta ^2+4 \beta +4\right) \epsilon ^2 \left(\frac{b}{r}\right)^{4/\beta }}{(\beta -2)^2 \beta ^2 r^4}+\frac{16 \Lambda  m \left(r^3-4 l^2 m\right)^2}{3 \left(2 l^2 m+r^3\right)^3}+\frac{16 \Lambda  m \left(4 l^2 m-r^3\right)}{3 \left(2 l^2 m+r^3\right)^2}+\frac{24 \Lambda ^2}{9}  \ .
\end{eqnarray}
Let's discuss different ranges for the parameters
\begin{enumerate}
    \item The first observation is $ \beta\neq 2$.
    \item For $\beta = -1$, the Kretschmann scalar remains finite throughout the spacetime and admits well-defined limits at both the origin and spatial infinity. Specifically, one finds:
\begin{equation}
\lim_{r \rightarrow 0} \mathcal{K} = \frac{8 \left(l^2 - b^2 \left(\Lambda l^2 + 3\right)\right)^2}{3 b^4 l^4}\;\;\;\text{and}\;\;\; \lim_{r \rightarrow \infty} \mathcal{K} = \frac{8 \left(b^2 \Lambda - 1 \right)^2}{3 b^4} \ .
\end{equation}
These expressions confirm that the curvature remains regular in both the ultraviolet and infrared regimes, indicating the absence of curvature singularities for $\beta=-1$. But for $\beta<-1$ it diverges near $r\rightarrow 0$ so we should exclude these regions.
\item For the parameter ranges $0 < \beta < 2$ and $\beta > 2$, the Kretschmann scalar exhibits divergent behavior in the vicinity of the origin, indicating a curvature singularity at $r = 0$. Conversely, the scalar remains finite at large radial distances, asymptotically approaching a constant value. Specifically, we have:
\begin{equation}
\lim_{r \rightarrow 0} \mathcal{K} = \infty \;\;\; \text{and}\;\;\; \lim_{r \rightarrow \infty} \mathcal{K} = \frac{8 \Lambda^2}{3} \ .
\end{equation}
This behavior reflects the emergence of a singularity at short distances, while the spacetime tends toward a (anti-)de Sitter geometry at infinity, depending on the sign of $\Lambda$.  
\end{enumerate}
The analysis reveals that the incorporation of a string fluid component significantly modifies the regularity properties of the Hayward black hole geometry. In particular, the spacetime ceases to be regular for values of the parameter $\beta < -1$ and $\beta > 0$, as evidenced by the divergence of the Kretschmann scalar at the origin. Conversely, within the interval $-1 \leq \beta < 0$, the regularity of the original Hayward solution is retained, indicating that the influence of the string fluid preserves the non-singular character of the spacetime only within this specific parameter range.

For Hayward black holes surrounded by string fluids in AdS spacetime, the extended first law of black hole thermodynamics takes the form,
\begin{eqnarray}\label{first law}
    dM = TdS +VdP + \Phi dq
\end{eqnarray}
where $M$ can be interpreted as the enthalpy of the black hole, $T$ is the thermodynamic temperature of the black hole, $q$ is the charge of the black hole, $\Phi$ is the potential conjugate to the charge $q$ and $V$ is the thermodynamic volume conjugate to the pressure $P$. In the extended phase space thermodynamics, the pressure can be associated with the cosmological constant via the relation,
\begin{equation}\label{Cosmological constant}
    P = -\frac{\Lambda}{8 \pi}.
\end{equation}

By equating Eq.~\eqref{Metric f_r} to zero at the horizon radius $r_+$, the mass or equivalently the enthalpy of the black hole can be expressed as,
\begin{equation}
    M = \frac{\left(q^3+r_+^3\right) \left(\beta  \varepsilon  l^2 \left(\frac{b}{r_+}\right)^{2/\beta }+(\beta -2) \left(l^2+r_+^2\right)\right)}{2 (\beta -2) l^2 r_+^2} \ .
\end{equation}
The thermodynamic temperature of the black hole is computed as,
\begin{equation}\label{Temp.}
    T= \frac{\varepsilon  l^2 \left(\frac{b}{r_+}\right)^{2/\beta } \left((\beta -2) r_+^3-2 (\beta +1) q^3\right)+(\beta -2) \left(l^2 \left(r_+^3-2 q^3\right)+3 r_+^5\right)}{4 \pi  (\beta -2) l^2 r_+ \left(q^3+r_+^3\right)}
\end{equation}
Further, the entropy $S$ and volume $V$ can be computed respectively as, 
\begin{equation}\label{S}
    S=\int \frac{dM}{T} = \pi r_+^2-\frac{\pi q^3}{r_+}, \quad V= \frac{\partial M}{\partial P} = \frac{4}{3}\pi r_{+}^3.
\end{equation}
This shows a correction to the standard Bekenstein-Hawking area law, due to the nonlinear electrodynamics and depends on charge $q$, and in the vanishing $q$ one gets back to the standard Bekenstein-Hawking area law. There are numerous other aspects of this solution discussed in recent years, e.g., particle dynamics and QPO's \cite{CALISKAN202499, Mustafa:2025lix, Ashraf:2025nvt, Mustafa:2025jco}, their thermodynamic stability \cite{doi:10.1142/S0219887825501439, Naseer:2025ghn, Naseer:2024tgt, Murtaza:2024ylz, Rasheed:2024aia}.

\section{Thermodynamic Curvature}

Ruppeiner geometry provides a powerful differential geometric framework to analyze the thermodynamic behavior of systems, including black holes. Introduced by Ruppeiner in 1979~\cite{Ruppeiner:1979bcp}, this approach constructs a Riemannian metric on the manifold of equilibrium states using the second derivatives of entropy with respect to extensive thermodynamic variables. The metric tensor, defined as the negative Hessian of the entropy function, leads to the line element
\begin{equation}\label{SRuppeiner}
ds_R^2 = - \frac{\partial^2 S}{\partial x^i \partial x^j} dx^i dx^j \ ,
\end{equation}
where $x^i$ denote the relevant extensive variables. In the context of fluctuation theory~\cite{Ruppeiner:1995zz}, the Ruppeiner distance between two points on this manifold encodes the likelihood of thermodynamic fluctuations between the corresponding states: larger distances correspond to lower fluctuation probabilities. Importantly, the scalar curvature derived from this metric provides insights into the nature of microscopic interactions, being zero for ideal systems, negative for systems with dominant attractive interactions, and positive when repulsive forces prevail.

In thermodynamic fluctuation theory~\cite{Ruppeiner:1995zz}, the number of accessible microstates of a system is exponentially related to its entropy, with the Boltzmann constant serving as the proportionality factor that connects macroscopic thermodynamic quantities to the underlying microscopic configurations. Consider a thermodynamic system $I_0$ in a state of equilibrium, encompassing a subsystem $I$ whose behavior is governed by fluctuations in a set of independent variables $x_i$ (with $i = 1, 2$). The likelihood $P(x_1, x_2)$ of the system residing within an infinitesimal vicinity of the state $(x_1, x_2)$ is intrinsically linked to the ensemble of accessible microstates. By virtue of the second law of thermodynamics, the equilibrium state is identified as the configuration that maximizes the entropy, $S = S_{\text{max}}$. Consequently, any perturbations in the variables $x_i$ may be regarded as subtle fluctuations oscillating around this point of maximal entropy. The probability distribution for such fluctuations~\cite{Ruppeiner:1995zz} near equilibrium is then given by
\begin{eqnarray}
 P(x_1, x_2) \propto e^{-\frac{1}{2} \Delta l^2} \ ,    
\end{eqnarray}
where $\Delta l^2$ is known as the thermodynamic line element and quantifies the distance between neighboring fluctuation states. This line element is defined as
\begin{eqnarray}\label{line element}
\Delta l^2 = -\frac{1}{k_B} \frac{\partial^2 S}{\partial x_i \partial x_j} \Delta x^i \Delta x^j \ .    
\end{eqnarray}
A smaller thermodynamic distance $\Delta l^2$ implies that fluctuations between nearby states are more likely. From this line element, one can compute the thermodynamic scalar curvature $R$, which serves as a powerful diagnostic tool for characterizing interactions within the system. The scalar curvature $R$, derived from this metric, has been extensively studied in various physical contexts, ranging from ideal and van der Waals gases to quantum systems, the Ising model, and black holes \cite{Mirza:2007ev, Quevedo:2007mj, Hendi:2015rja, Belhaj:2015uwa, Bhattacharya:2017hfj, Wei:2019uqg, Wei:2019yvs}. Current understanding suggests that positive $R$ indicates repulsive interactions, negative $R$ corresponds to attractive interactions, and vanishing $ R$ characterizes a non-interacting or balanced system. Importantly, $R$ often diverges at critical points associated with phase transitions, making it a valuable geometric indicator of underlying microscopic changes in the system.

This geometric interpretation becomes particularly insightful when applied to black hole thermodynamics. Recent developments suggest that black holes may possess microscopic constituents analogous to molecules in conventional thermodynamic systems~\cite{Wei:2015iwa, AR, Mann2019, Wei:2019yvs, SAdS}. The Ruppeiner curvature thus offers a window into the interactions among these hypothesized microstructures. For instance, curved Ruppeiner metrics imply the presence of interactions, with the sign of the curvature indicating their nature. Historically, the utility of Ruppeiner geometry in black hole physics was first recognized in studies involving the BTZ black hole~\cite{Cai:1998ep}, and later extended to a variety of spacetimes, including Reissner–Nordström, Kerr, and AdS black holes~\cite{Shen:2005nu}. These analyses confirmed that divergences in the scalar curvature align with known thermodynamic phase transitions, such as those characterized by Davies~\cite{Davies:1978mf}. Subsequent research has further explored the application of this formalism to a wide class of gravitational systems, demonstrating its robustness in revealing thermodynamic behavior~\cite{Dehyadegari, Sheykhi, Miao3, Chen}. Beyond traditional thermodynamic geometries, recently the role of contact geometry in modeling the full thermodynamic phase space~\cite{hermann,mrugala1996} has been emphasized. This framework treats equilibrium processes as trajectories on Legendre submanifolds—subspaces of the contact manifold that correspond to thermodynamic states. Within this setting, the Ruppeiner metric emerges as a restriction of a more general contact-compatible Riemannian structure defined over the full phase space~\cite{contactBH}. While there are ongoing discussions surrounding Legendre-invariant formulations and their implications for the uniqueness and physical interpretation of thermodynamic metrics~\cite{Quevedo:2006xk, Zhang:2015ova, Hendi:2015xya}. In this work, we focus on the physical insights provided by Ruppeiner curvature and its interplay between thermodynamic geometry and black hole microstructure, which remains a fertile ground for uncovering quantum gravitational phenomena.

For cases like Schwarzschild black holes, thermodynamic geometry cannot be constructed in the absence of additional parameters describing the black hole (such as charge or angular momentum), since all of the thermodynamic quantities rely solely on the horizon radius. However, a variable cosmological constant offers the extra parameter in extended thermodynamics, making it possible to analyze neutral black holes as well, albeit with the need to construct a new line element. For instance, the potential might be chosen as enthalpy, $H= H(S,P)$, in the extended thermodynamic setup, where the fluctuation coordinates are just $S$ and $P$. A detailed computation of line elements in the $(S,P)$-planes can be found in \cite{Ghosh:2023khd}. An internal energy representation, i.e. the $(S,V)$-plane is not suitable for many black hole systems where the fluctuation coordinates $S$ and $V$ are not independent. The choice of metric to use in a particular situation depends on the features of the system one is interested in studying. The curvature scalar may not be equivalent if it is computed employing different ensembles. However, this should not be a major concern because all that is required is that one chooses the appropriate fluctuation coordinates for the ensemble one wants to work in \cite{Bravetti:2012dn}.  One key observation regarding the line elements is that, the curvature (as well as the line elements) will generally have divergences at the point where the response functions (such as specific heats etc..) diverge and this is expected based on the phase structure of the theory. It is well known that some of the divergences observed in the specific heats do not appear in the Ruppeiner curvature\cite{Ghosh:2023khd,Bravetti:2012dn} depending on the parameterization and fluctuation coordinates employed. This led to a search for alternate line elements, with varying degrees of success\cite{Dolan:2015xta,Singh:2020tkf,Singh:2023hit}. There are other line elements one can construct, which effectively capture phase transition and critical points. These are particularly helpful in situations where the response functions diverge trivially, as will occur in the scenario examined in our work, and that needs to be eliminated by hand. In such cases, a novel normalized scalar curvature can be used, which was pioneered in \cite{Wei:2019uqg}. This important metric which can be constructed in the extended thermodynamics, turns out to effectively capture the phase transition and critical points of charged black holes as well as van der Waals systems. There are of course a plethora of other line elements that one can envisage and construct, depending on the thermodynamic ensemble, with each containing varying information of the system. For a detailed computation of line elements in various planes using different potentials, one can follow the appendix~\ref{Appendix B} of this paper. As the list is quite long and beyond
the scope of this work, we refer the reader to the review in reference \cite{Wei:2019yvs,Ghosh:2023khd}, where a thorough treatment of different line elements and their physical aspects has also been
discussed in detail.

\quad Further, using the line element in Eq. (\ref{line element}), it is straightforward to construct the Ruppeiner metric  in the  \((T,V)\)-plane in the extended thermodynamics  which can effectively capture the phase transition and critical points of charged black holes as,
\begin{equation}\label{metric}
dl_{R_{N}}^2 = \frac{1}{T}\bigg(\frac{\partial P}{\partial V}\bigg)_TdV^2 + \frac{C_V}{T^2}dT^2 
\end{equation}
The given metric is diagonal but becomes singular when the specific heat at constant volume, \(C_V\), vanishes--a condition satisfied by all static AdS black holes. By rescaling the metric or its scalar curvature by \(C_V\), this singularity can be resolved, resulting in a normalized thermodynamic scalar curvature that is frequently written as $R_N$ in the literature\cite{Singh:2023hit,Singh:2023ufh, Singh:2020tkf, Singh:2025ueu}.

We emphasize the geometric consistency of all stated thermodynamic line elements and discuss them in appendix~\ref{Appendix B}. The metrics are connected by conformal factors as one transitions from one potential (and its natural variables) to another under a Legendre transform. For instance, the $(T,V)$ metric we will employ here is exactly the same as the $(S,P)$ form and is just the enthalpy-based metric expressed in $(T,V)$ variables. There are no inconsistencies. For the Hayward black hole with string fluid, the $(T,V)$ coordinate system is a plausible and practically suitable choice. It is derived directly from the traditional method of constructing thermodynamic metrics for the extended ensemble using the Helmholtz free energy (or enthalpy). Thus we shall, in this work, study the Ruppeiner geometry on the $(T,V)$-plane as discussed in the following subsection.


\subsection{Thermodynamic Curvature of Hayward Black Holes
Surrounded by String Fluids}
This section discusses the thermodynamic curvature of Hayward black holes surrounded by string fluids. We start by writing the temperature as in Eq.~\eqref{Temp.} in the terms of pressure as \begin{eqnarray}\label{Temperature}
   T = \frac{\epsilon  \left(\frac{b}{r_+}\right)^{2/\beta } \left((\beta -2) r_{+}^3-2 (\beta +1) q^3\right)-(\beta -2) \left(-8 \pi  P r_{+}^5+2 q^3-r_{+}^3\right)}{4 \pi  (\beta -2) r_+ \left(q^3+r_{+}^3\right)}
\end{eqnarray}
To see the behavior of the thermodynamic temperature of the black hole with horizon radius, we plot Eq.~\eqref{Temperature}.
\begin{figure}[h!]
    \begin{center}
        \includegraphics[scale=.50]{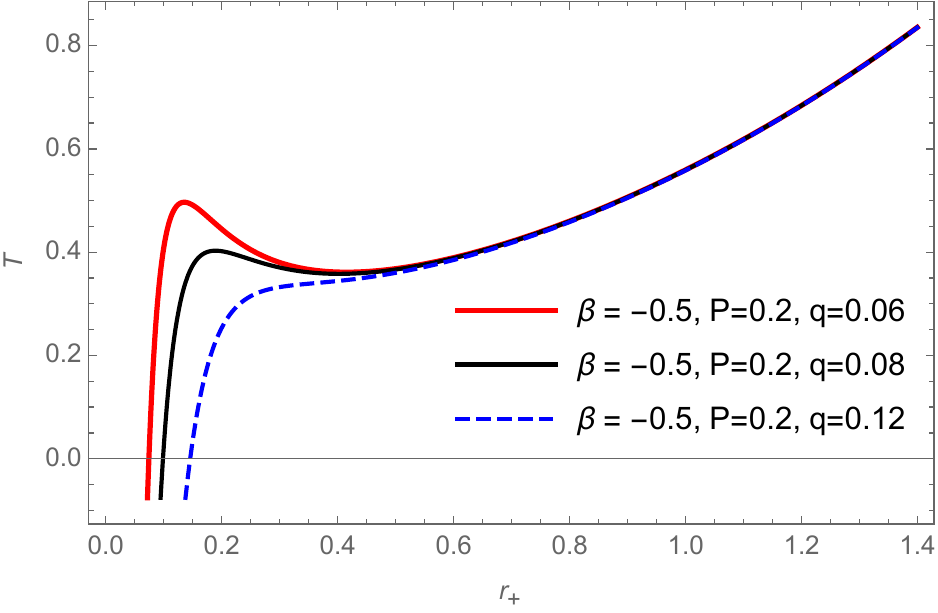} 
    \end{center}
    \caption{The behavior of black hole temperature $T$ with horizon radius $r_{+}$ for fixed values of $\epsilon=1, b=1, \beta=-0.5, P=0.2$ and charge $q$.}
    \label{T_vs_r}
\end{figure}

\begin{figure}[h!]
    \begin{center}
        \includegraphics[scale=.50]{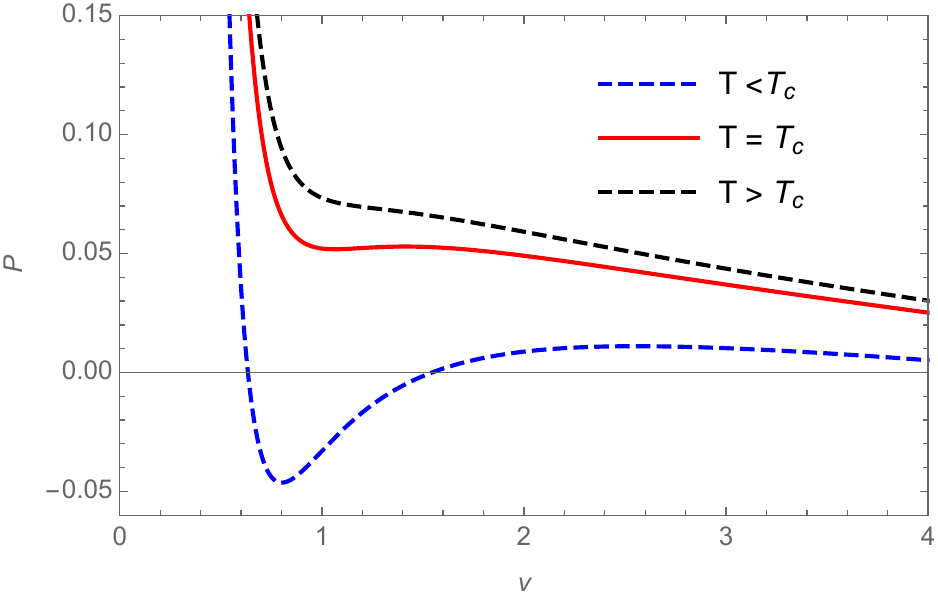} 
    \end{center}
    \caption{The behavior of thermodynamic pressure $P$ with specific volume $v$ for fixed values of $q=0.2, \beta=-0.5, \epsilon=1, b=1$ and thermodynamic temperature $T$.}
    \label{P_vs_v}
\end{figure}

From Fig.~(\ref{T_vs_r}), one can note that the Hawking temperature is positive. For Hayward black holes with string fluids, the temperature-horizon radius plot exhibits non-monotonic behavior, suggesting some phase transitions from small to large black holes as the black hole charge $q$ increases. This behavior is similar to that of van der Waals fluids, demonstrating a rich thermodynamic structure that is controlled by the parameters $q, \epsilon, \beta$ and thermodynamic pressure $P$ of the black hole. Now, solving Eq.~\eqref{Temperature} for thermodynamic pressure $P$ and taking the specific volume as $v=2r_+$, we can obtain the equation of state as
\begin{equation}
P =\frac{T \left(8 q^3+v^3\right)}{v^4}-\frac{v^3-16 q^3}{2 \pi  v^5} -\frac{2^{\frac{2}{\beta }-1} \epsilon  \left(\frac{b}{v}\right)^{2/\beta } \left((\beta -2) v^3-16 (\beta +1) q^3\right)}{\pi  (\beta -2) v^5}
\label{Eos}
\end{equation}
The behavior of thermodynamic pressure $P$ with specific volume $v$, i.e. the $P-v$ diagram, is displayed in Fig.~\ref {P_vs_v}.

The $(P-v)$ diagram displays temperature-dependent isotherms. At temperatures below the critical temperature $T_c$, the isotherms indicate van der Waals-like behavior, signifying unstable regions. The boundary of thermodynamic instability (negative compressibility) is marked by the spinodal curve, which can be found from the extrema of $P(v)$ \cite{Nascimento:2024maj}. The critical pressure decreases with increasing black hole charge $q$, suggesting suppressed phase transitions. The $P–v$ diagram offers information on stable, metastable, and unstable configurations, highlighting the rich phase structure of regular AdS black holes. Together with the obtained equation of state, the resulting phase diagram affirms that Hayward AdS black holes with string fluids experience phase transitions that are comparable to those observed in standard thermodynamic systems.
\par
\begin{figure}[h!]
    \begin{center}
        \includegraphics[scale=.60]{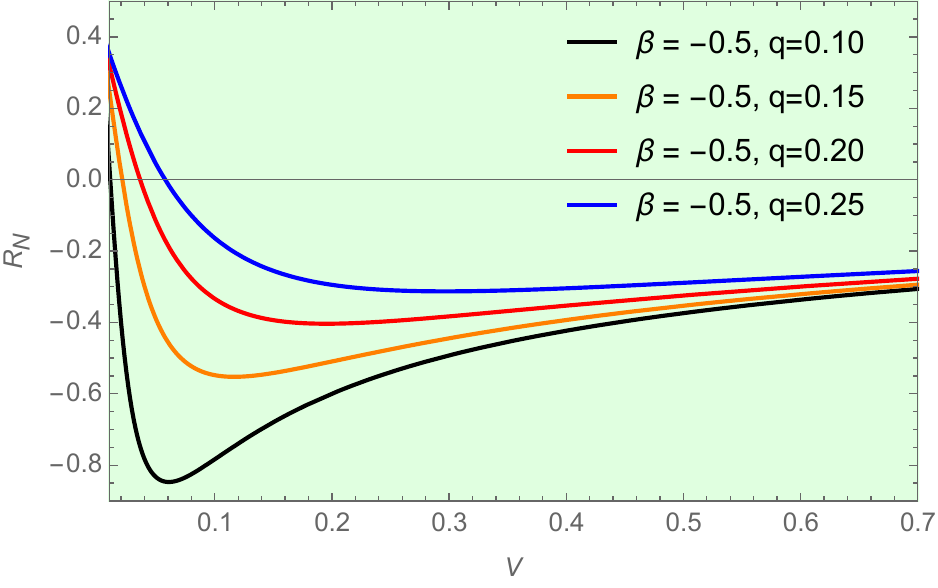} 
    \end{center}
    \caption{Thermodynamic curvature $R_N$ with black hole volume $V$ for fixed values of $\epsilon=1, b=1, \beta=-0.5, T=1.2$ and charge $q$}
    \label{RN_vs_V}
\end{figure}

 
Now, utilizing Eq. (\ref{Temperature}), Eq. (\ref{Eos}) (expressing it in terms of thermodynamic volume $V$ using Eq. (\ref{S}) instead of specific volume $v$) and Eq. (\ref{metric}), we can compute and write the analytical expressions for the Ruppeiner curvature in $(T,V)$-plane as,
\begin{eqnarray}
    R_{N}= \frac{R_A}{R_B}
\end{eqnarray}
where,
\begin{eqnarray}
&& R_A = \Bigg[2^{\frac{4}{3 \beta}} \pi^{\frac{2}{3 \beta}} (\beta + 1) \epsilon \left(\frac{b}{\sqrt[3]{V}}\right)^{\frac{2}{\beta}} \left(4 \pi (5 \beta + 2) q^3 - 3 (\beta - 2) V \right) + 3^{\frac{2}{3 \beta}} (\beta - 2) \beta \left(20 \pi q^3 - 3 V \right)\Bigg] \nonumber \\
 &&\times \Bigg\{2^{\frac{4}{3 \beta}} \pi^{\frac{2}{3 \beta}} (\beta + 1) \epsilon \left(\frac{b}{\sqrt[3]{V}}\right)^{\frac{2}{\beta}} \left(4 \pi (5 \beta + 2) q^3 - 3 (\beta - 2) V \right)+ 3^{\frac{2}{3 \beta}} (\beta - 2) \beta \Big(4 \pi q^3 \big(8 \sqrt[3]{6} \pi^{\frac{2}{3}} T \sqrt[3]{V} \nonumber\\
 && \;\;\;\;\;\;\;\;\;\;\;\;\;\;\;\;\;\;\;\;\;\;\;\;\;\;\;\;\;\;\;\;\;\;\;\;\;\;\;\;\;\;\;\;\;\;\;\;\;\;\;\;\;\;\;\;\;\;\; + 5 \big) + 6 \sqrt[3]{6} \pi^{\frac{2}{3}} T V^{\frac{4}{3}} - 3 V \Big)\Bigg\} \\
&& R_B = 2 \Bigg[2^{\frac{4}{3 \beta}} \pi^{\frac{2}{3 \beta}} (\beta + 1) \epsilon \left(\frac{b}{\sqrt[3]{V}}\right)^{\frac{2}{\beta}} \left(4 \pi (5 \beta + 2) q^3 - 3 (\beta - 2) V \right) + 3^{\frac{2}{3 \beta}} (\beta - 2) \beta \Big( 4 \pi q^3 \big(4 \sqrt[3]{6} \pi^{\frac{2}{3}} T \sqrt[3]{V} \nonumber \\
&&\;\;\;\;\;\;\;\;\;\;\;\;\;\;\;\;\;\;\;\;\;\;\;\;\;\;\;\;\;\;\;\;\;\;\;\;\;\;\;\;\;\;\;\;\;\;\;\;\;\;\;\;\;\;\;\;\;\;\; + 5 \big) + 3 V \big(\sqrt[3]{6} \pi^{\frac{2}{3}} T \sqrt[3]{V} - 1 \big) \Big)\Bigg]^2
\end{eqnarray}
The behavior of the thermodynamic curvature with volume of the black hole is displayed in Fig. \ref{RN_vs_V}.

The nature of microscopic interactions in the black hole system can be determined by plotting the normalized Ruppeiner curvature $R_N$ with thermodynamic volume $V$ for Hayward AdS black holes with string fluids. Across all curves, $R_N$ is predominantly negative in the major volume regime, indicating the dominance of attractive interactions between the underlying microscopic constituents. This is consistent with the general interpretation in Ruppeiner geometry indicating that a negative scalar curvature corresponds to attractive interactions, similar to the behavior of bosonic systems or charged AdS black holes. $R_N$ is typically less negative or approaches zero in a very small volume regime and in some cases is slightly positive for higher charges $q$. A positive $R_N$ suggests repulsive interactions, resembling fermionic systems. Thus, for large $q$, the system begins to exhibit short-range repulsion at small volume regimes, indicating a shift in microstructure behavior with increasing charge. Here, the qualitative behavior of $R_N$ is comparable to that of regular AdS black holes and Reissner–Nordström AdS black holes, where similar attractive interactions dominate but with brief regions of repulsive nature at small volume regimes or high charges. These observations support the interpretation of black holes as thermodynamic systems with nontrivial microscopic structure.

The scalar curvature $R$ is known to depend on the choice of ensemble. For the same black hole, different coordinate choices (canonical, grand-canonical, and extended) result in different $R$. In the grand-canonical ensemble, for instance, $R$ changes from negative to positive when the temperature is lowered, but for charged AdS black holes, $R>0$ (highly repulsive) is found for all configurations in the canonical ensemble. Each metric is nevertheless valid within its ensemble: one simply fixes different variables. Thus, although the coordinate choice changes the numerical curvature, it reflects the correct physics under the imposed constraints. By using $(T,V)$ in this instance, we are treating temperature and volume as independent variables and dealing with the free energy (or enthalpy). This is suitable for the extended ensemble of ``black hole chemistry" that we take into consideration. It is important to stress that there is no conflict in the use of $(T,V)$ in this context because all thermodynamic measurements are mutually consistent (connected via Legendre transformations and the factor $1/T$). The key point is that which metric one uses must match the ensemble. Finally, we note the consistency with other recent studies \cite{Cong:2021jgb}. For example, \cite{Cong:2021jgb} explicitly observe that no $(P,V)$ criticality arises for charged AdS black holes, which is precisely what we can also find when trying a $(P,V)$ formulation. Instead, they demonstrate that phase transitions occur only in temperature-related ensembles (such as fixed charge or fixed chemical potential), which supports our choice of $T$ as one of the coordinates. Further, the study by Cong et al. introduces an extended thermodynamic framework that incorporates the central charge and its conjugate chemical potential, providing a novel perspective on black hole phase transitions through the lens of holography. In contrast, our focus on the $(T,V)$ coordinate system offers a detailed understanding of the interactions among the microstructures of black holes. Both approaches are valuable, each shedding light on different aspects of black hole thermodynamics. Furthermore, the same $(T,V)$ structure is also necessary for our thermodynamic topology study, which introduces winding numbers and topological charge $W$. Consistent with $R$ changing sign and diverging at the correct points, we find that the number and type of topological charges vary throughout the transition and are dependent on the black hole parameters. Thus, a self-consistent and physically relevant thermodynamic geometry is obtained by selecting $(T,V)$ as the fluctuation coordinates. The Ruppeiner curvature singularities in these coordinates closely match the known small-to-large black hole transition points, and the curvature sign clearly indicates between regimes of repulsive and attractive microstructure. Other coordinate choices, such as $(S,P)$ or $(P,V)$, either make it more difficult to determine $R$ or fail to indicate the true instabilities. As a result, we think that the $(T,V)$ representation is both convenient for our study and natural (through the Helmholtz potential), as supported by the literature. This decision guarantees that our geometric and topological probes accurately capture the phase behavior of Hayward black holes with string fluids without affecting any theory.



\section{Thermodynamic Topology}

Recent advancements in black hole thermodynamics have underscored the importance of topological and geometric frameworks in probing critical phenomena and phase structures. Among these, Duan’s $\phi$-mapping topological current theory \cite{a19, a20} has emerged as a robust method for identifying and classifying phase transitions through the study of vector field singularities in the thermodynamic parameter space. Unlike conventional approaches based solely on the analysis of thermodynamic potentials, this method exploits the correspondence between phase transition points and the zeros of a carefully constructed vector field $\boldsymbol{\phi}$, interpreted as topological defects. The singularities of $\boldsymbol{\phi}$ act as sources of a conserved topological current, defined through a generalized Jacobian structure. This current admits a delta-function-like support on the critical points of the system and is mathematically expressed in terms of the Jacobian determinant of the mapping from the thermodynamic manifold to the order parameter space. The associated topological invariant, denoted by $W$, is computed via a local decomposition involving the Hopf index and the Brouwer degree, which encapsulate the multiplicity and orientation of vector field windings near its singularities. This invariant provides a global measure of phase structure and is sensitive to topological bifurcations, thereby capturing subtle thermodynamic instabilities and metastable configurations.

To implement this framework, the Euclidean action formalism is invoked, wherein thermodynamic equilibrium is enforced by compactifying the temporal direction with period $\tau = T^{-1}$, where $T$ is the Hawking temperature \cite{a19, a20}. A generalized free energy function is then introduced as
\begin{equation}\label{T1}
\mathscr{F} = M - \frac{S}{\tau},
\end{equation}
with $M$ representing the ADM mass and $S$ the black hole entropy. This quantity is stationary under equilibrium configurations and forms the basis for constructing the $\boldsymbol{\phi}$-field through partial derivatives with respect to the relevant variables, typically the event horizon radius $r_+$ and an auxiliary angular coordinate $\Theta$. Explicitly, the vector field is defined by
\begin{equation}\label{T2}
\phi = \left( \frac{\partial \mathscr{F}}{\partial r_+}, -\cot \Theta \csc \Theta \right),
\end{equation}
with the second component ensuring that $\phi$ diverges at $\Theta = 0, \pi$, effectively confining the analysis within a bounded topological domain.

The topological current $j^\mu$ associated with the field $\phi$ is expressed as
\begin{equation}\label{T3}
j^\mu = \frac{1}{2\pi} \epsilon^{\mu\nu\rho} \epsilon^{ab} \partial_\nu n^a \partial_\rho n^b, \quad n^a = \frac{\phi^a}{|\phi|},
\end{equation}
where $\mu, \nu, \rho = 0,1,2$ and $a, b \in \{1, 2\}$. This current is identically conserved, i.e., $\partial_\mu j^\mu = 0$, and its support is localized at the zero points of $\phi$. The total topological charge, or winding number, is then computed via surface integration:
\begin{equation}\label{T4}
W = \int_{\Sigma} j^0 , d^2x = \sum_{i=1}^{n} \zeta_i \eta_i \ ,
\end{equation}
where $\zeta_i$ is the Hopf index and $\eta_i$ is the sign of the Jacobian determinant evaluated at the $i$-th zero of $\phi$. The summation yields the net topological invariant associated with the thermodynamic system. A positive topological charge corresponds to a stable phase with well-defined thermodynamic behavior, while a negative or vanishing value typically signals instability, such as saddle-node bifurcations or first-order transitions. This methodology has been successfully applied to a wide range of black hole models, including those embedded in asymptotically AdS/dS spacetimes, higher-derivative gravity theories, and setups involving non-linear electrodynamics and scalar hair~\cite{ 22a, 23a, 25a, 27a, 31a, 33a, 34a, 38', 38a, 38b, 38c, 43a, 44a, 44e, 44g, 44h, 44i, 44j, 44k, 44l, 44m, Sekhmani:2024vsu}. The $\phi$-mapping framework thus offers a universal and mathematically rigorous approach for classifying black hole phase structures in terms of topological and geometric invariants, furnishing a deeper understanding of gravitational thermodynamics from a global perspective.


\subsection{Thermodynamic Topology of Hayward Black Holes
Surrounded by String Fluids}

This subsection examines the thermodynamic topology of black holes. Building upon Eq.~\eqref{T1}, the generalized Helmholtz free energy for Hyward-AdS black holes is expressed as,
\begin{equation}\label{T5}
\mathscr{F}= \frac{1}{2 r_+^2} \left[\frac{\left(q^3+r_+^3\right) \left(\beta  \varepsilon  l^2 \left(\frac{b}{r_+}\right)^{2/\beta }+(\beta -2) \left(l^2+r_+^2\right)\right)}{(\beta -2) l^2}-\frac{2 \pi  r_+ \left(r_+^3-2 q^3\right)}{\tau } \right]\ .
\end{equation}
Moreover, utilizing Eq.~(\ref{T2}), the components of the vector field \( \phi \) are explicitly determined as,
\begin{equation}\label{T6}
\begin{split}
&\phi^{r_+}= \frac{1}{2 r_+^3} \left[2 q^3 \left(-\frac{(\beta +1) \varepsilon  \left(\frac{b}{r_+}\right)^{2/\beta }}{\beta -2}-1\right)+r_+^3 \left(\varepsilon  \left(\frac{b}{r_+}\right)^{2/\beta }+1\right)+\frac{3 r_+^5}{l^2}-\frac{4 \pi  r_+ \left(q^3+r_+^3\right)}{\tau }\right] \ .
\end{split}
\end{equation}
By equating them to zero, we can easily compute the $\tau$ as
\begin{equation}\label{T7}
\begin{split}
\tau =\frac{4 \pi  (\beta -2) l^2 r_+ \left(q^3+r_+^3\right)}{\varepsilon  l^2 \left(\frac{b}{r_+}\right)^{2/\beta } \left((\beta -2) r_+^3-2 (\beta +1) q^3\right)+(\beta -2) \left(l^2 \left(r_+^3-2 q^3\right)+3 r_+^5\right)} \ .
\end{split}
\end{equation}

\begin{figure}[]
 \begin{center}
 \subfigure[]{
 \includegraphics[height=4.1cm,width=4.1cm]{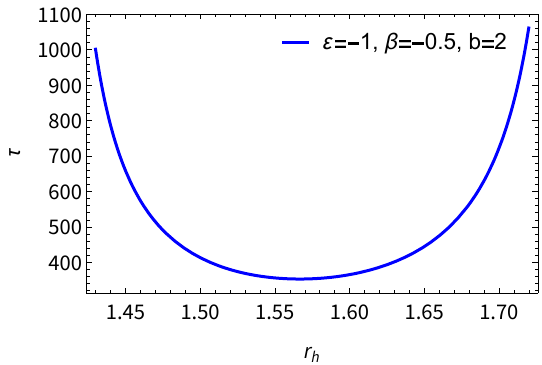}
 \label{100a}}
 \subfigure[]{
 \includegraphics[height=4.1cm,width=4.1cm]{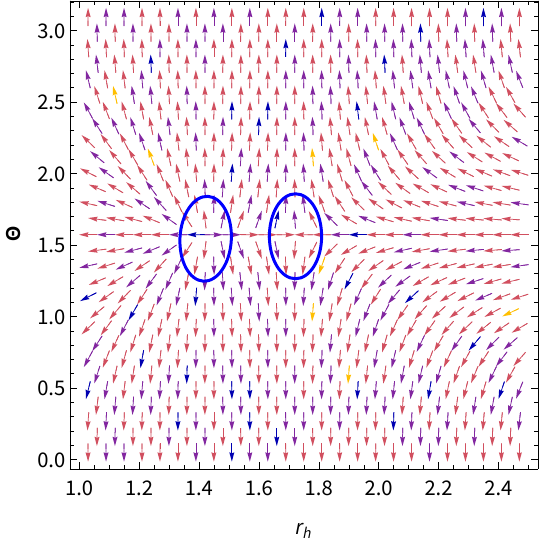}
 \label{100b}}
 \subfigure[]{
 \includegraphics[height=4.1cm,width=4.1cm]{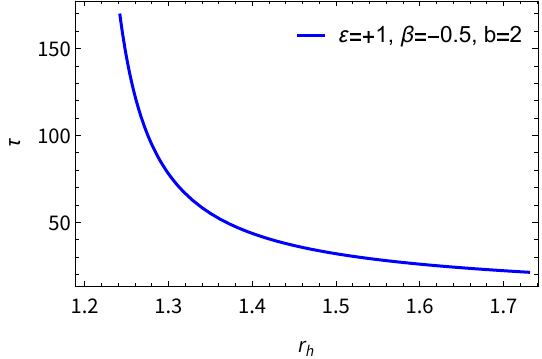}
 \label{100c}}
 \subfigure[]{
 \includegraphics[height=4.1cm,width=4.1cm]{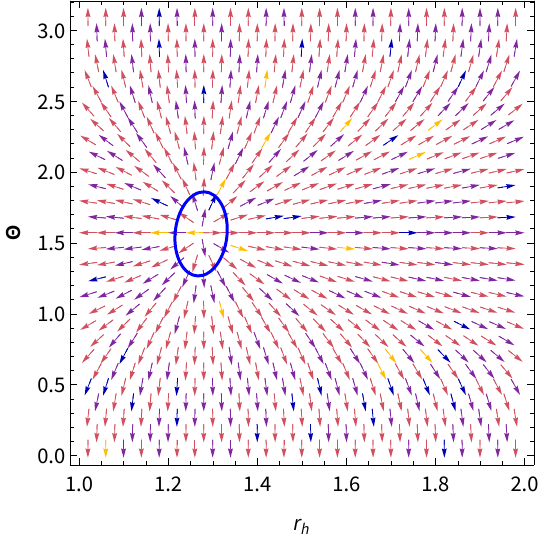}
 \label{100d}}
 \subfigure[]{
 \includegraphics[height=4.1cm,width=4.1cm]{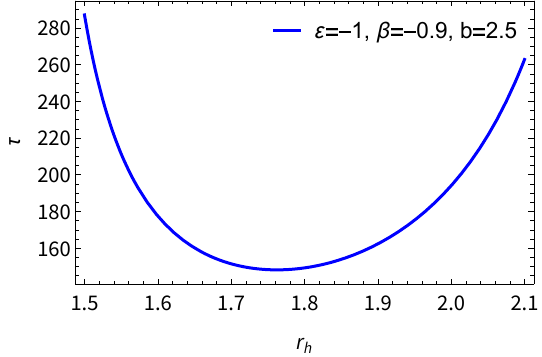}
 \label{100e}}
 \subfigure[]{
 \includegraphics[height=4.1cm,width=4.1cm]{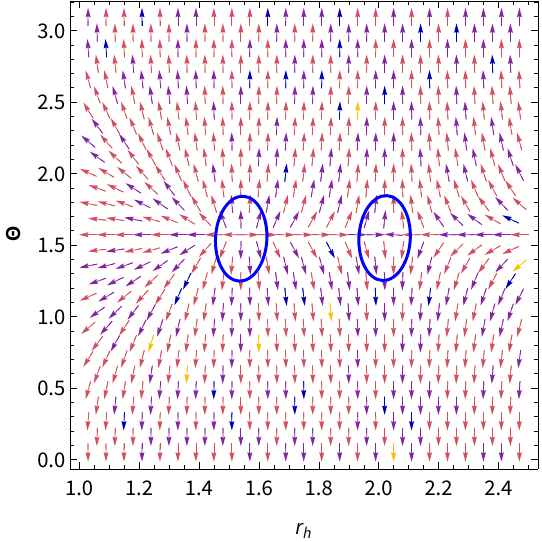}
 \label{100f}}
 \subfigure[]{
 \includegraphics[height=4.1cm,width=4.1cm]{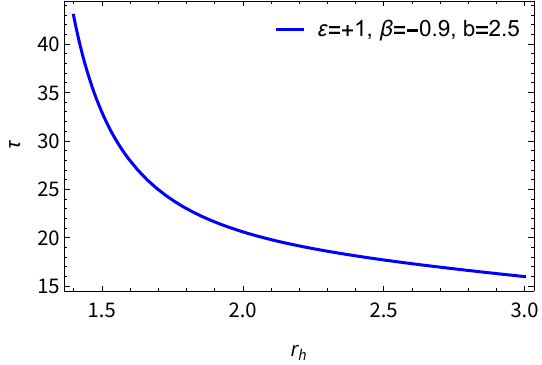}
 \label{100g}}
 \subfigure[]{
 \includegraphics[height=4.1cm,width=4.1cm]{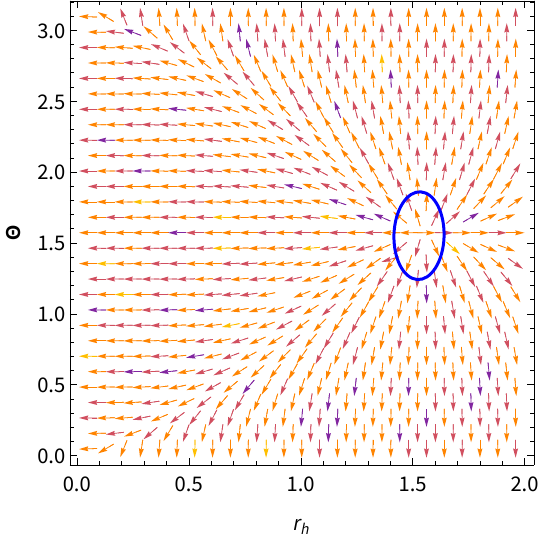}
 \label{100h}}
 \subfigure[]{
 \includegraphics[height=4.1cm,width=4.1cm]{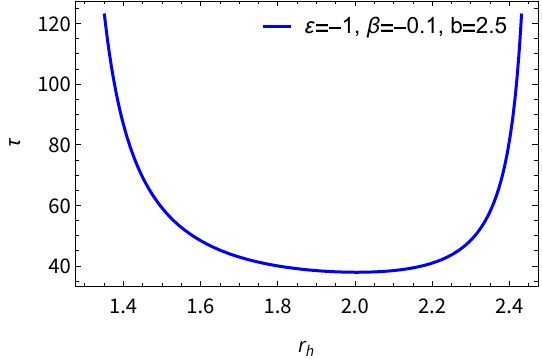}
 \label{100i}}
 \subfigure[]{
 \includegraphics[height=4.1cm,width=4.1cm]{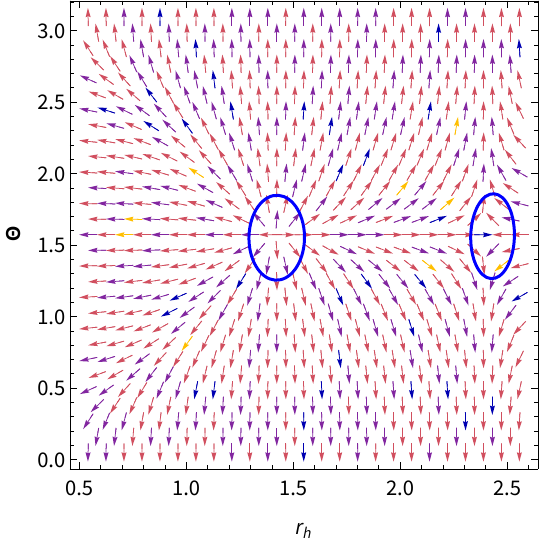}
 \label{100j}}
 \subfigure[]{
 \includegraphics[height=4.1cm,width=4.1cm]{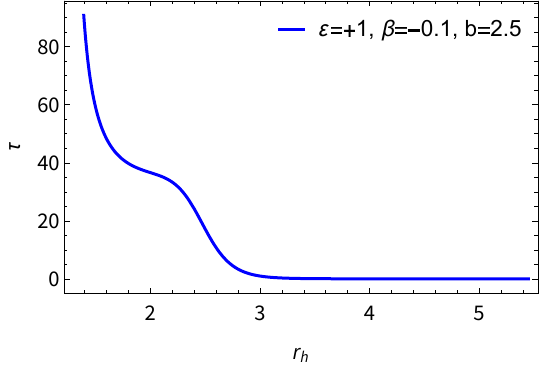}
 \label{100k}}
 \subfigure[]{
 \includegraphics[height=4.1cm,width=4.1cm]{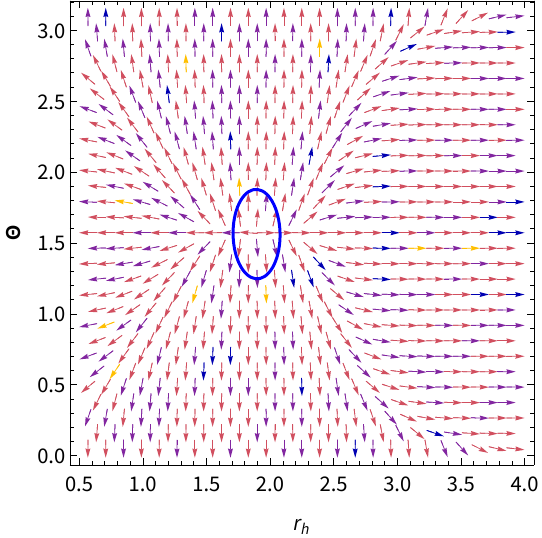}
 \label{100l}}
 \subfigure[]{
 \includegraphics[height=4.1cm,width=4.1cm]{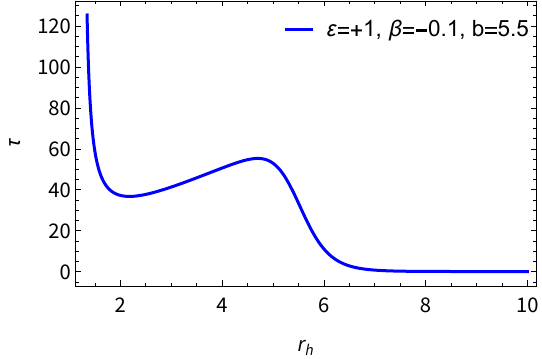}
 \label{100m}}
 \subfigure[]{
 \includegraphics[height=4.1cm,width=4.1cm]{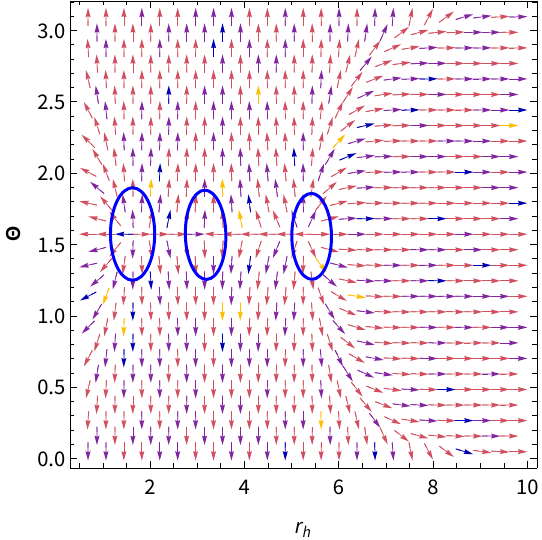}
 \label{100n}}
 \caption{The \((\tau \text{ vs. } r_+)\) diagram, illustrating variations in free parameters for Hayward-AdS black holes, is displayed in Figs.~(\ref{100a}), (\ref{100c}), (\ref{100e}), (\ref{100g}), (\ref{100i}), (\ref{100k}), and (\ref{100m}). Additionally, the normal vector field \( n \) in the \((r_+, \Theta)\) plane is represented, with Zero Points (ZPs) appearing at specific coordinates \( (r_+, \Theta) \). These ZPs are associated with parameter values \( b = 2, 2.5, 5.5 \), \( \epsilon = -1, +1 \), and \( \beta = -0.1, -0.5, -0.9 \).}
 \label{m1}
\end{center}
 \end{figure}

Now, we explore the thermodynamic topology of these black holes, with accompanying figures illustrating the distribution of topological charges. The normalized field lines, depicted in Fig.~(\ref{m1}), provide insight into the properties of topological charges within the system. Figures~(\ref{100b}), (\ref{100f}), and (\ref{100j}) highlight the existence of two zero points at specific coordinates \((r_+, \Theta)\), corresponding to the parameter value \(\epsilon = -1\) and various values of \( \beta \) and \( b \). These zero points are integral to understanding the local topological characteristics of the black hole. In contrast, Figures~(\ref{100d}), (\ref{100h}), and (\ref{100l}) display a single zero point for \(\epsilon = +1\), indicating a shift in topological behavior under certain parameter adjustments.

The zero points, which serve as representations of topological charges, are enclosed within blue contour loops. Their ordering and spatial distribution are directly influenced by variations in parameters \(\epsilon\) and \(b\). A significant feature emerging from these results is that when \(\epsilon = -1\), two topological charges, \( (\omega = +1, -1) \), manifest, leading to a total topological charge of \( W = 0 \) and when \(\epsilon =+1\) one topological charge, \( (\omega = +1) \), manifest, leading to a total topological charge of \( W = +1 \). This phenomenon is demonstrated in Figures~(\ref{m1}).

To assess the stability of the black holes, an analysis based on the number of coils within the normalized field lines is conducted. By modulating the parameter \(\epsilon\), the classification of topological charge structures undergoes a transition, as evidenced in Figures~(\ref{100d}), (\ref{100h}), and (\ref{100l}). Under these circumstances, a single topological charge, \( \omega = +1 \), emerges, resulting in a total topological charge of \( W = +1 \).

Additionally, when \( \epsilon = +1 \) and the parameter \( b \) increases, the number of topological charges also increases. However, despite this increase, the classification remains consistent, maintaining a total topological charge of \( W = +1 \) (see Fig.~(\ref{100n})). Thus, Hayward-AdS black holes exhibit distinct topological classes that are critically dependent on the parameters \(\epsilon\) and \(b\), which play a fundamental role in determining both the classification of black holes and the number of associated topological charges.

To further refine the topological framework, we analyze the free energy function as a scalar quantity defined within the two-dimensional parameter space \((r_h, \Theta)\). The corresponding vector field \( \phi \) is formulated so that its zero points coincide with the extrema of the free energy function. The rotational characteristics of the field lines around these zero points—indicative of whether they correspond to maxima or minima—serve as the basis for systematically assigning topological charges \cite{a19}. A comparative examination of fundamental black hole solutions, including Schwarzschild and Reissner-Nordström configurations, reveals distinct patterns in their topological charge values:
\begin{enumerate}
    \item Schwarzschild black hole: \( W = -1 \)  
           \item Reissner-Nordström black hole: \( W = 0 \)  
               \item AdS-Reissner-Nordström black hole: \( W = +1 \) 
\end{enumerate}
To classify black holes based on their topological charges, a systematic exploration of varying free parameters within the model is required. By examining different parameter ranges and evaluating the corresponding topological charges, a more refined classification scheme emerges. Specifically, when \( \epsilon = -1 \), the resulting black hole exhibits characteristics analogous to the Reissner-Nordström solution \( W = 0 \) . Conversely, for \( \epsilon = +1 \), the black hole demonstrates behavior akin to the AdS-Reissner-Nordström configuration \( W = +1 \), thereby facilitating a more structured categorization within the broader framework of conventional black hole models \cite{a19}.
These classical solutions provide essential insights into black hole thermodynamics, enabling a systematic classification of black hole structures and their associated thermodynamic properties. Within the scope of this study, our findings align with established theoretical predictions, as confirmed in \cite{a19}. This topological methodology not only reinforces theoretical expectations but also establishes a structured approach for assessing black hole stability and phase transitions, thereby extending its applicability to gravitational thermodynamics and astrophysical investigations.

\section{Conclusion}

In this paper, we have developed and investigated a class of regular black hole solutions that employ a surrounding fluid of strings and a negative cosmological constant to extend the original Hayward geometry. The geometric and thermodynamic features of the spacetime structure are enhanced by significant modifications that are caused by the inclusion of string fluids. We derived the corresponding metric solutions and rigorously examined their regularity by evaluating the Kretschmann scalar. According to our investigation, the spacetime is only regular for the parameter range $-1 \le \beta <0$. On the other hand, the Kretschmann scalar diverges at the origin of $\beta=2$, which has been associated with perfect fluid dark matter theories, suggesting a central singularity. One of the significant aspects of our work is the analysis of equation of state for the Hayward black hole with string fluids, which exhibits behavior analogous to that of van der Waals fluids. We examined the dependence of the thermodynamic pressure $P$ on significant parameters such as the charge $q$ and the string fluids parameter $b$ by expressing it as a function of the specific volume $v=2r_{+}$. The resulting $ P-v$ diagram, which is depicted in Figure (\ref{P_vs_v}), clearly illustrates the existence of oscillating isotherms and inflection points below the critical temperature $T_C$, which are characteristics of first-order phase transitions. Hayward AdS black holes with string fluids exhibit phase transitions similar to those in classical thermodynamic systems, as validated by this thermodynamic behavior. Critical phenomena are greatly influenced by the parameters $q$ and $b$. These results emphasize the intricate and modified thermodynamic framework that string fluids provide.

\quad In addition to the above study, we used empirical methods from thermodynamic geometry, namely Ruppeiner geometry, to investigate the stability structure and microscopic interactions of these string fluid-modified Hayward AdS black holes. Ruppeiner thermodynamic geometry can be utilized to investigate the microstructure of these black holes. The microscopic nature of the interactions can be determined by curvature:the normalized Ruppeiner curvature $R_N$. Repulsion is indicated by $R_N>0$, while effectively attracting ``black hole molecules" are indicated by $R_N<0$. The majority of parameter spaces for Hayward AdS black holes with string fluids have a negative $R_N$, which is dominated by attractive interactions. Nonetheless, a small region of repulsive interaction, akin to that observed in charged AdS holes, is observed in the small black hole branch. The Reissner–Nordström–AdS context is, in fact, replicated in the qualitative pattern of $R_N$: small black holes exhibit $R_N>0$ (repulsive), whereas $R_N$ at the critical point indicates a second-order transition. Significantly, the normalized curvature $R_N$ diverges precisely at the phase transition, establishing a one-to-one correspondence between $R_N$ singularities and heat capacity divergences. As a result, $R_N$ offers an efficient diagnostic: it not only indicates criticality in the extended thermodynamic phase space, but also provides microscopic details on the repulsive or attractive interactions between black hole microstructures. This approach enabled a deeper understanding of phase behavior and critical phenomena within the extended thermodynamic framework, highlighting the impact of string fluids on black hole microstructure.

\quad We carried out an in-depth analysis of the thermodynamic topology of Hayward black holes surrounded by string fluids. The normalized field lines illustrate the underlying topological charge structure of the system. When the parameter $\epsilon$ is set to $-1$, the system exhibits two distinct zero points in the $(r_+, \Theta)$ space. These correspond to topological charges of $\omega = +1$ and $\omega = -1$, resulting in a net topological charge $W = 0$. Conversely, when $\epsilon = +1$, only a single zero point with charge $\omega = +1$ is observed, giving a total charge of $W = +1$. These zero points, which are enclosed by closed contour loops, shift in position and number depending on the values of $\epsilon$ and the parameter $b$. A stability analysis based on the winding behavior of the field lines reveals a clear topological transition driven by the sign of $\epsilon$. Furthermore, increasing the value of $b$ when $\epsilon = +1$ leads to the appearance of additional charges, yet the total charge remains $W = +1$. This demonstrates that the topological classification of Hayward-AdS black holes is highly sensitive to both $\epsilon$ and $b$, as these parameters determine the multiplicity of charges and the overall topological structure.

\quad In conclusion, the paradigm of black hole chemistry is extended to a regular, singularity-free context by Hayward AdS black holes with string fluid. Their Ruppeiner geometry and van der Waals-like thermodynamics support the idea that black hole microphysics exhibits universal patterns: the sign and divergence of $R_N$ reveal the underlying intermolecular interactions in direct analogy with well-known black hole and fluid systems, while the string fluids and charge tune the critical behavior. The thermodynamic topology of Hayward-AdS black holes is governed by parameters $\epsilon$ and $b$, which control the number and nature of topological charges in the phase space. The system undergoes a topological transition $\epsilon$ changes, with the total charge $W$ serving as a global invariant characterizing the black hole’s phase structure. This topological analysis reveals critical information about the phase structure and stability of Hayward-AdS black holes. The number and type of topological charges reflect the presence of phase transitions, while the total charge $W$ serves as a global invariant indicating the overall thermodynamic stability and configuration of the system. Our knowledge of black hole phase structure and microstructure in extended thermodynamics is enhanced by these results, which imply that regular black holes can be regarded as well-behaved areas for quantum gravity and AdS/CFT research.

The key takeaways of the current study. We have developed and analyzed a class of regular black hole solutions modified by a surrounding fluid of strings and a negative cosmological constant, extending the original Hayward geometry. The inclusion of string fluids introduces profound modifications to the spacetime structure, influencing both its geometric and thermodynamic properties.
\begin{enumerate}
 \item The spacetime remains regular for $-1 \leq \beta <0$, for other values of $\beta$, the Kretschmann scalar diverges, indicating a central singularity linked to perfect fluid dark matter theories.
 \item The black hole's thermodynamic pressure $P$ as a function of specific volume mirrors van der Waals fluid behavior. The resulting $P-v$ diagram exhibits oscillatory isotherms and first-order phase transitions, significantly influenced by charge $(q)$ and string fluid parameter $(b)$.
 \item The application of Ruppeiner geometry elucidates the microscopic interactions. Negative $R_N$ values indicate predominant attractive microstructures, while a small domain of repulsive interaction emerges, similar to charged AdS black holes.
 \item The divergence of $R_N$ at phase transitions precisely coincides with heat capacity singularities, providing a direct diagnostic for criticality in the extended thermodynamic space.
 \item The topological charge distribution depends on $\epsilon$ and $b$.
\begin{itemize}
    \item  For $\epsilon = -1$, the resulting black hole exhibits characteristics analogous to the Reissner-Nordstrom solution with total topological charge $W=0$
    \item For $\epsilon = +1$,  the black hole demonstrates behavior akin to the AdS-Reissner-Nordstrom configuration with total topological charge $W = +1$.
    \item Increasing $b$ affects charge distribution but maintains the total charge. A topological transition occurs when $\epsilon$ changes, highlighting its role in stability and phase behavior.
\end{itemize}
\end{enumerate}
Hayward AdS black holes modified by string fluids present a robust thermodynamic framework where microstructures, stability, and phase transitions align with classical van der Waals fluids and charged AdS systems. The thermodynamic topology analysis further provides a global invariant $(W)$ that dictates black hole stability. These insights enhance our understanding of black hole phase structure and microphysics, offering valuable implications for quantum gravity and AdS/CFT research. This extended framework positions regular black holes as viable candidates for exploring quantum gravity models and holographic correspondences. The interplay between string fluids, charge, and topology reveals universal thermodynamic signatures, paving the way for deeper explorations in black hole chemistry and microstructural interactions.

An important direction for future work would be to explore potential observational signatures of string fluid black holes. This includes studying modifications to black hole shadows, shifts in quasi-normal mode spectra, and possible imprints on gravitational wave signals from black hole mergers. We consider the effect of string fluids on
\begin{itemize}
    \item \textbf{Shadows:} The string fluid may modify the photon sphere radius and hence the angular size and shape of the black hole shadow, potentially making it distinguishable from standard Schwarzschild or Kerr profiles.
    \item \textbf{Quasi-Normal Modes:} Since the effective potential for perturbations depends on the background metric, the presence of a string fluid could alter the QNM spectrum, which could in principle be constrained via gravitational wave observations from black hole mergers.
    \item \textbf{Gravitational Waves:} The ringdown phase of gravitational wave signals is particularly sensitive to the black hole’s near-horizon geometry. Therefore, deviations due to a string fluid could leave detectable imprints in gravitational wave templates, especially with future precision observations.
Such analyses would help connect the theoretical framework of string fluids with astrophysical observations and could provide a means to constrain or detect their presence in the strong gravity regime.
\end{itemize}
We will get back to these in our next work.

\appendix
 
\section{Energy-momentum tensor for the string fluid}
The energy momentum tensor(EMT) for the cloud of particles is characterized by only the kinetical part 
\begin{eqnarray}
    \mathrm{K}^{\mu \nu} = \rho_p u^\mu u^\nu \ .
\end{eqnarray}
The tensor $K_{\mu\nu}$ represents the kinetical part of the EMT. The kinetical part of the EMT refers to the component arising solely from the motion of particles. Also, $u^\mu$ is the normalized flux density, and $\rho_p$ is the fluid's rest energy density, and subscript $p$ represents the particles. The full EMT can be written as 
\begin{eqnarray}\label{EMT for particle}
    T^{\mu \nu} \rho_p u^\mu u^\nu +S^{\mu\nu} \ ,
\end{eqnarray}
where $S^{\mu\nu}$ represents the stress-energy tensor. For $u^\mu$ to be the flux density, it should be an eigenvector of $T^{\mu\nu}$, and for that, $S^{\mu\nu}u_\nu=0$ must be satisfied. So, the simplest choice for $d^{\mu \nu}$ is 
\begin{equation}
    S^{\mu\nu}=ph^{\mu\nu}\;\;\; \text{where} \;\;\;h^{\mu\nu}=g^{\mu\nu}-u^\mu u^\nu \ .
\end{equation}
The Eq.~\eqref{EMT for particle} describes the usual EMT for a perfect fluid. 
Let's make a similar discussion for strings. The kinetical part for the strings is 
\begin{eqnarray}
    \mathrm{K}^{\mu\nu} = \rho\;\sqrt{-\gamma}\;\Sigma^{\mu \lambda}\Sigma^\nu_\lambda/(-\gamma) \ .
\end{eqnarray}
It is easy to verify $ \mathrm{K}^{\mu\nu} \Sigma^\lambda_\mu = \rho \sqrt{-\gamma}\Sigma^{\mu \lambda}$. Now, we can write the EMT for string is 
\begin{eqnarray}
    T^{\mu \nu} = \rho \sqrt{-\gamma}\;\Sigma^{\mu \lambda}\Sigma^\nu_\lambda/(-\gamma) +S^{\mu\nu} \ .
 \end{eqnarray}
 Again the condition imposed on $S^{\mu \nu}$ is $S^{\mu \nu}\Sigma^\lambda_\nu=0$. The choice for  $S^{\mu \nu}$ is
 \begin{eqnarray}
      S^{\mu \nu} = pH^{\mu\nu} \;\;\;\text{where}\;\;\; H^{\mu\nu} = g^{\mu\nu}-\Sigma^{\alpha \mu}\Sigma^\nu_\alpha \ .
 \end{eqnarray}
 Finally, the EMT for the fluid of strings is \begin{eqnarray}
     T^{\mu\nu} = \left(p + \sqrt{-\gamma}\rho\right) \frac{\Sigma^{\mu\beta} \Sigma_\beta^{;\nu}}{-\gamma} + p g^{\mu\nu} \ .
 \end{eqnarray}
 It exactly matched with Eq.~\eqref{EMT for String}. For the case of a static, spherically symmetric spacetime, the symmetry constraints restrict the bivector $\Sigma^{\mu\nu}$ to possess only two non-vanishing components: $\Sigma^{tr}$ and $\Sigma^{\theta\phi}$. Consequently, due to the underlying symmetries, the energy-momentum tensor for a perfect string fluid simplifies such that the temporal and radial components are equal, i.e., $T^t_t = T^r_r$, while the angular components satisfy $T^\theta_\theta = T^\phi_\phi = p$. The gravitational field becomes uniquely determined once an equation of state is imposed, specifying the relationship between the temporal component and the angular components of the energy-momentum tensor. We adopt the assumption that $T^t_t$ is proportional to $T^\theta_\theta$. Accordingly, the energy-momentum tensor takes the form:
\begin{eqnarray}
    T^t_t = T^r_r = - \beta  T^\theta_\theta \ ,
\end{eqnarray} 
where $\beta$ serves as a dimensionless parameter encoding the anisotropic properties of the energy-momentum tensor, distinguishing between the radial and tangential pressures of the fluid. The form of $T^{\mu \nu}$ is chosen as in Eq.~\eqref{eq:1.17.1}. Then the differential equation we have to solve to get the metric function \eqref{eq:f_r} is
\begin{equation}
f''+\frac{f'}{r}-\frac{f}{r^2}+\Lambda -\frac{12l^2m^2r^2}{r^2(r^3+2l^2m)^2}-48\frac{(r^3-l^2m)l^2m^2r^2}{r^2(r^3+2l^2m)^3}+\frac{\epsilon}{r^2}\left(\frac{\beta+2}{\beta}\right)\left(\frac{b}{r}\right)^{2/\beta}=0 \ .
\end{equation}


\section{Derivation of the Ruppeiner line elements on different planes}\label{Appendix B}
In order to derive the line element of the Ruppeiner metric with independent coordinates $S$ and $P$, let us start out with the generic expression:
\begin{equation}
dl^2_R=-g_{\mu \nu}dx^\mu dx^\nu
\end{equation}
where $g_{\mu \nu}=\partial \mu \partial \nu S.$
Writing out $dz_\mu =g_{\mu \nu}dx^\nu$ one gets,
\begin{equation}
dl^2_R=-dz_\mu dx^\mu
\end{equation}
Therefore, we must have
\begin{equation}
z_\mu =\frac{\partial S}{\partial x^\mu.}
\end{equation}
Now, the first law is given as,
\begin{equation}\label{first law}
dM=TdS+VdP
\end{equation}
Now, from the first law given in eq.(\ref{first law}), one can write,
\begin{equation}
dS-\frac{1}{T}dM + \frac{V}{T}dP=0
\end{equation}
which means that $z_1 = 1/T$ and $z_2 = -V/T$ whereas $x^1 = M$ and $x^2 = P$ such that $ds = z_\mu dx^\mu.$ Then, with these identifications,
\begin{eqnarray}
dz_1=-\frac{dT}{T^2},\\ dz_2=\frac{V}{T^2}dT-\frac{1}{T}dV.
\end{eqnarray}
The line element given has the form $dl^2_{R} = -dz_{1} dx^1 - dz_{2} dx^2$
can now be written as,
\begin{eqnarray}
dl^2_R &=& -\bigg(-\frac{dT}{T^2} \bigg)dM-\bigg(\frac{V}{T^2}dT-\frac{1}{T}dP \bigg)dV, \\
&=&\frac{dT}{T^2}dM-\frac{V}{T^2}dPdT+\frac{1}{T}dPdV, \\
&=&\frac{dT}{T^2}(TdS+VdP)-\frac{V}{T^2}dPdT+\frac{1}{T}dPdV
\end{eqnarray}
Hence, from the first law it reduces to
\begin{equation}\label{Line element}
dl^2_R=\frac{dSdT}{T}+\frac{dPdV}{T}
\end{equation}
\subsubsection{$(S,P)$-plane}
At this stage, if one considers $S$ and $P$
to be independent coordinates such that,
\begin{eqnarray}
T= T(S,P), V= V(S,P)
\end{eqnarray}
and consequently,
\begin{equation}
\Delta T=\bigg(\frac{\partial T}{\partial S}\bigg)\Delta S+\bigg(\frac{\partial T}{\partial P}\bigg)\Delta P
\end{equation}
\begin{equation}
\Delta V=\bigg(\frac{\partial V}{\partial S}\bigg)\Delta S+\bigg(\frac{\partial V}{\partial P}\bigg)\Delta P
\end{equation}
Substituting them in eq.(\ref{Line element}) and using the Maxwell's relations, 
\begin{eqnarray}
   \bigg(\frac{\partial T}{\partial P}\bigg)_{S}= \bigg(\frac{\partial V}{\partial S}\bigg)_{P}
\end{eqnarray}
we can write,
\begin{equation}\label{SP metric}
dl^2_R=\frac{1}{C_p}\Delta S^2+\frac{2}{T}\bigg(\frac{\partial T}{\partial P}\bigg)\Delta S \Delta P+\frac{1}{T}\bigg(\frac{\partial V}{\partial P}\bigg)\Delta P^2
\end{equation}
\subsubsection{$(T,\mu)$-plane}
In order to derive the line element of the Ruppeiner metric with independent coordinates $T$ and $\mu$, one must follow the same procedure as above.
Taking $z_1 = 1/T$ and $z_2 = -\mu/T$ whereas $x^1 = M$ and $x^2 = Q$ such that $ds = z_\mu dx^\mu.$ Then, with these identifications, the first law it reduces to
\begin{equation}\label{Line element2}
dl^2_R=\frac{dT dS}{T}-\frac{d\mu dQ}{T}
\end{equation}
If one considers $T$ and $\mu$
to be independent coordinates such that,
\begin{eqnarray}
S=S(T,\mu), Q=Q(T,\mu)
\end{eqnarray}
then on taking the derivatives,
\begin{equation}
\Delta S=\bigg(\frac{\partial S}{\partial T}\bigg)\Delta T+\bigg(\frac{\partial S}{\partial \mu}\bigg)\Delta \mu
\end{equation}
\begin{equation}
\Delta Q=\bigg(\frac{\partial Q}{\partial T}\bigg)\Delta T+\bigg(\frac{\partial Q}{\partial \mu}\bigg)\Delta \mu
\end{equation}
Substituting them in eq.(\ref{Line element}) one finally gets the line element in the fixed $Q$ ensemble as,
\begin{equation}\label{Tmu metric}
dl^2_R=\frac{C_\mu}{T^2}\Delta T^2+\frac{2}{T}\bigg(\frac{\partial S}{\partial \mu}\bigg)\Delta T \Delta \mu-\frac{1}{T}\bigg(\frac{\partial P}{\partial \mu}\bigg)\Delta \mu^2
\end{equation}
For fixed potential, $\Omega=U-\mu Q, \quad dU =TdS -Qd \mu$, one can write the corresponding line element in the $(T, \mu)$-plane as,
\begin{equation}
  dl^2_R=\frac{C_\mu}{T^2}\Delta T^2-\frac{1}{T}\bigg(\frac{\partial Q}{\partial \mu}\bigg)\Delta \mu^2
\end{equation}
\subsubsection{$(T,V)$-plane}
If one considers $S$ and $P$
to be independent coordinates such that,
\begin{eqnarray}
S=S(T,V), P=P(T,V)
\end{eqnarray}
further, one gets,
\begin{equation}
\Delta S=\bigg(\frac{\partial S}{\partial T}\bigg)\Delta T+\bigg(\frac{\partial S}{\partial V}\bigg)\Delta V
\end{equation}
\begin{equation}
\Delta P=\bigg(\frac{\partial P}{\partial T}\bigg)\Delta T+\bigg(\frac{\partial P}{\partial V}\bigg)\Delta V
\end{equation}
Here, Helmholtz free energy is a function of $T$ and$V$ and it follows the relation(see \cite{Wei:2019yvs} for detailed computation of different line elements),
\begin{equation}
  \bigg(\frac{\partial S}{\partial V}\bigg)= -\bigg(\frac{\partial P}{\partial T}\bigg)   
\end{equation}
Substituting them in eq.(\ref{Line element}) and following exactly the similar computation as above, one can obtain the corresponding line element in the $(T,V)$-plane as,
\begin{equation}
  dl^2_R=\frac{C_V}{T^2}\Delta T^2+\frac{1}{T}\bigg(\frac{\partial P}{\partial V}\bigg)\Delta V^2
\end{equation}
A large number of thermodynamic line elements, each suited to a specific ensemble or thermodynamic potential, can be constructed in the context of black hole thermodynamic geometry. Numerous aspects of the black hole's thermodynamic behavior, including stability, phase transitions, and critical occurrences, are captured by these metrics. Importantly, we stress that all these entities are geometrically consistent: Legendre transformations mandate conformal transformations that link the thermodynamic metrics. Because each thermodynamic potential has its own set of natural variables, the accompanying metrics are conformally equivalent as one moves between them, preserving the underlying geometric structure that the Legendre framework imposes.


\section*{Acknowledgments}

A.A. is financially supported by the Institute's postdoctoral fellowship at IITK. A.S. would like to thank CSIR-HRDG for the financial support received as a Postdoctoral Research Associate working under Project No. 03WS(003)/2023-24/EMR-II/ASPIRE.

\bibliographystyle{main}
\bibliography{main}

\providecommand{\href}[2]{#2}\begingroup\raggedright\begin{thebibliography}{100}

\bibitem{Bardeen:1973gs}
J.M.~Bardeen, B.~Carter and S.W.~Hawking, \emph{{The Four laws of black hole
  mechanics}}, \href{https://doi.org/10.1007/BF01645742}{\emph{Commun. Math.
  Phys.} {\bfseries 31} (1973) 161}.

\bibitem{Bekenstein:1973ur}
J.D.~Bekenstein, \emph{{Black holes and entropy}},
  \href{https://doi.org/10.1103/PhysRevD.7.2333}{\emph{Phys. Rev. D} {\bfseries
  7} (1973) 2333}.

\bibitem{Hawking:1975vcx}
S.W.~Hawking, \emph{{Particle Creation by Black Holes}},
  \href{https://doi.org/10.1007/BF02345020}{\emph{Commun. Math. Phys.}
  {\bfseries 43} (1975) 199}.

\bibitem{Maldacena:1997re}
J.M.~Maldacena, \emph{{The Large $N$ limit of superconformal field theories and
  supergravity}}, \href{https://doi.org/10.4310/ATMP.1998.v2.n2.a1}{\emph{Adv.
  Theor. Math. Phys.} {\bfseries 2} (1998) 231}
  [\href{https://arxiv.org/abs/hep-th/9711200}{{\ttfamily hep-th/9711200}}].

\bibitem{Hawking:1982dh}
S.W.~Hawking and D.N.~Page, \emph{{Thermodynamics of Black Holes in anti-De
  Sitter Space}}, \href{https://doi.org/10.1007/BF01208266}{\emph{Commun. Math.
  Phys.} {\bfseries 87} (1983) 577}.

\bibitem{Cong:2021jgb}
W.~Cong, D.~Kubiznak, R.B.~Mann and M.R.~Visser, \emph{{Holographic CFT phase
  transitions and criticality for charged AdS black holes}},
  \href{https://doi.org/10.1007/JHEP08(2022)174}{\emph{JHEP} {\bfseries 08}
  (2022) 174} [\href{https://arxiv.org/abs/2112.14848}{{\ttfamily
  2112.14848}}].

\bibitem{Cui:2021qpu}
Y.-z.~Cui, W.~Xu and B.~Zhu, \emph{{Hawking-Page transition with reentrance and
  triple point in Gauss-Bonnet gravity}},
  \href{https://doi.org/10.1103/PhysRevD.107.044048}{\emph{Phys. Rev. D}
  {\bfseries 107} (2023) 044048}
  [\href{https://arxiv.org/abs/2106.13942}{{\ttfamily 2106.13942}}].

\bibitem{Bambi:2013ufa}
C.~Bambi and L.~Modesto, \emph{{Rotating regular black holes}},
  \href{https://doi.org/10.1016/j.physletb.2013.03.025}{\emph{Phys. Lett. B}
  {\bfseries 721} (2013) 329}
  [\href{https://arxiv.org/abs/1302.6075}{{\ttfamily 1302.6075}}].

\bibitem{Barca:2023shv}
G.~Barca and G.~Montani, \emph{{Non-singular gravitational collapse through
  modified Heisenberg algebra}},
  \href{https://doi.org/10.1140/epjc/s10052-024-12564-5}{\emph{Eur. Phys. J. C}
  {\bfseries 84} (2024) 261}
  [\href{https://arxiv.org/abs/2309.09767}{{\ttfamily 2309.09767}}].

\bibitem{Dymnikova:1992ux}
I.~Dymnikova, \emph{Vacuum nonsingular black hole},
  \href{https://doi.org/10.1007/BF00760226}{\emph{Gen. Rel. Grav.} {\bfseries
  24} (1992) 235}.

\bibitem{PhysRevD.20.1294}
P.S.~Letelier, \emph{Clouds of strings in general relativity},
  \href{https://doi.org/10.1103/PhysRevD.20.1294}{\emph{Phys. Rev. D}
  {\bfseries 20} (1979) 1294}.

\bibitem{Letelier}
P.S.~Letelier, \emph{Fluids of strings in general relativity},
  \href{https://doi.org/10.1007/BF02755096}{\emph{Il Nuovo Cimento B
  (1971-1996)} {\bfseries 63} (1981) 519}.

\bibitem{Soleng1995}
H.H.~Soleng{\emph{General Relativity and Gravitation} {\bfseries 27} (1995)
  367}.

\bibitem{Soleng1994}
H.H.~Soleng, \emph{Correction to einstein's perihelion precession formula from
  a traceless, anisotropic vacuum energy},
  \href{https://doi.org/10.1007/BF02105150}{\emph{General Relativity and
  Gravitation} {\bfseries 26} (1994) 149}.

\bibitem{Ruppeiner1995}
G.~Ruppeiner, \emph{Riemannian geometry in thermodynamic fluctuation theory},
  \href{https://doi.org/10.1103/RevModPhys.67.605}{\emph{Rev. Mod. Phys.}
  {\bfseries 67} (1995) 605}.

\bibitem{Ruppeiner2010}
G.~Ruppeiner, \emph{Thermodynamic curvature measures interactions},
  \href{https://doi.org/10.1119/1.3488300}{\emph{Am. J. Phys.} {\bfseries 78}
  (2010) 1170}.

\bibitem{Zhang2020}
M.~Zhang, R.~Zhao and M.-S.~Ma, \emph{Thermodynamic topology of black hole
  phase transition in massive gravity},
  \href{https://doi.org/10.1016/j.physletb.2019.135109}{\emph{Phys. Lett. B}
  {\bfseries 800} (2020) 135109}.

\bibitem{Banerjee2022}
R.~Banerjee and D.~Sarkar, \emph{Thermodynamic geometry and topological
  classification of phase transitions in black holes},
  \href{https://doi.org/10.1103/PhysRevD.106.084021}{\emph{Phys. Rev. D}
  {\bfseries 106} (2022) 084021}.

\bibitem{Wei2019}
S.~Wei and Y.-X.~Liu, \emph{Topology of black hole thermodynamics},
  \href{https://doi.org/10.1103/PhysRevD.100.064006}{\emph{Phys. Rev. D}
  {\bfseries 100} (2019) 064006}.

\bibitem{Yerra2022}
P.K.~Yerra and C.~Bhamidipati, \emph{Topology of phase transitions in ads black
  holes}, \href{https://doi.org/10.1103/PhysRevD.106.084027}{\emph{Phys. Rev.
  D} {\bfseries 106} (2022) 084027}.

\bibitem{hayward2006formation}
S.A.~Hayward, \emph{Formation and evaporation of nonsingular black holes},
  {\emph{Physical review letters} {\bfseries 96} (2006) 031103}.

\bibitem{nascimento2024some}
F.F.~Nascimento, V.B.~Bezerra and J.M.~Toledo, \emph{Some remarks on hayward
  black hole surrounded by a cloud of strings}, {\emph{Annals of Physics}
  {\bfseries 460} (2024) 169548}.

\bibitem{molina2021thermodynamics}
M.~Molina and J.~Villanueva, \emph{On the thermodynamics of the hayward black
  hole}, {\emph{Classical and Quantum Gravity} {\bfseries 38} (2021) 105002}.

\bibitem{bronnikov2001regular}
K.A.~Bronnikov, \emph{Regular magnetic black holes and monopoles from nonlinear
  electrodynamics}, {\emph{Physical Review D} {\bfseries 63} (2001) 044005}.

\bibitem{fan2016construction}
Z.-Y.~Fan and X.~Wang, \emph{Construction of regular black holes in general
  relativity}, {\emph{Physical Review D} {\bfseries 94} (2016) 124027}.

\bibitem{bronnikov2017comment}
K.A.~Bronnikov, \emph{Comment on “construction of regular black holes in
  general relativity”}, {\emph{Physical Review D} {\bfseries 96} (2017)
  128501}.

\bibitem{toshmatov2018comment}
B.~Toshmatov, Z.~Stuchl{\'\i}k and B.~Ahmedov, \emph{Comment on “construction
  of regular black holes in general relativity”}, {\emph{Physical Review D}
  {\bfseries 98} (2018) 028501}.

\bibitem{soleng1995dark}
H.H.~Soleng, \emph{Dark matter and non-newtonian gravity from general
  relativity coupled to a fluid of strings}, {\emph{General Relativity and
  Gravitation} {\bfseries 27} (1995) 367}.

\bibitem{dymnikova1992vacuum}
I.~Dymnikova, \emph{Vacuum nonsingular black hole}, {\emph{General relativity
  and gravitation} {\bfseries 24} (1992) 235}.

\bibitem{soleng1994correction}
H.H.~Soleng, \emph{Correction to einstein's perihelion precession formula from
  a traceless, anisotropic vacuum energy},
  \href{https://doi.org/https://doi.org/10.1007/BF02105150}{\emph{General
  relativity and gravitation} {\bfseries 26} (1994) 149}.

\bibitem{salgado2003simple}
M.~Salgado, \emph{A simple theorem to generate exact black-hole solutions},
  {\emph{Classical and Quantum Gravity} {\bfseries 20} (2003) 4551}.

\bibitem{giambo2002anisotropic}
R.~Giambo, \emph{Anisotropic generalizations of de sitter spacetime},
  {\emph{Classical and Quantum Gravity} {\bfseries 19} (2002) 4399}.

\bibitem{dymnikova2002cosmological}
I.~Dymnikova, \emph{The cosmological term as a source of mass},
  {\emph{Classical and Quantum Gravity} {\bfseries 19} (2002) 725}.

\bibitem{toledo2020black}
J.M.~Toledo and V.B.~Bezerra, \emph{Black holes with a fluid of strings},
  {\emph{Annals of Physics} {\bfseries 423} (2020) 168349}.

\bibitem{Nascimento:2024maj}
F.F.~Nascimento, P.H.~Morais, J.M.~Toledo and V.B.~Bezerra, \emph{{Some remarks
  on Bardeen-AdS black hole surrounded by a fluid of strings}},
  \href{https://doi.org/10.1007/s10714-024-03268-y}{\emph{Gen. Rel. Grav.}
  {\bfseries 56} (2024) 86} [\href{https://arxiv.org/abs/2412.00552}{{\ttfamily
  2412.00552}}].

\bibitem{CALISKAN202499}
A.~Caliskan, G.~Mustafa, T.~Naseer, S.~Maurya, E.~Güdekli, S.~Murodov et~al.,
  \emph{{Particle dynamics with trajectories and epicyclic oscillations around
  a piece-wise black hole immersed in dark matter}},
  \href{https://doi.org/https://doi.org/10.1016/j.jheap.2024.09.005}{\emph{Journal
  of High Energy Astrophysics} {\bfseries 44} (2024) 99}.

\bibitem{Mustafa:2025lix}
G.~Mustafa, S.K.~Maurya, T.~Naseer, A.~Cilli, E.~G\"udekli, A.~Abd-Elmonem
  et~al., \emph{{Particle motion and QPOs around Euler-Heisenberg black hole
  immersed in cold dark matter halo}},
  \href{https://doi.org/10.1016/j.nuclphysb.2025.116812}{\emph{Nucl. Phys. B}
  {\bfseries 1012} (2025) 116812}.

\bibitem{Ashraf:2025nvt}
A.~Ashraf, T.~Naseer, H.~Chaudhary, A.~Bouzenada, F.~Atamurotov, B.~\c{C}il
  et~al., \emph{{Observational constraints on QPOs with orbital motion around
  charged non-commutative Schwarzschild black hole surrounded by perfect fluid
  dark matter}},
  \href{https://doi.org/10.1016/j.nuclphysb.2025.116873}{\emph{Nucl. Phys. B}
  {\bfseries 1014} (2025) 116873}.

\bibitem{Mustafa:2025jco}
G.~Mustafa, A.~Ditta, T.~Naseer, S.K.~Maurya, P.~Channuie, A.A.~Ibraheem
  et~al., \emph{{Circular motion, QPOs testing, emission energy and thermal
  fluctuations around a non-singular hairy Bardeen black hole}},
  \href{https://doi.org/10.1140/epjc/s10052-025-14235-5}{\emph{Eur. Phys. J. C}
  {\bfseries 85} (2025) 575}.

\bibitem{doi:10.1142/S0219887825501439}
T.~Naseer, \emph{{Thermodynamic stability and fluid behavior in charged Hayward
  black hole}},
  \href{https://doi.org/10.1142/S0219887825501439}{\emph{International Journal
  of Geometric Methods in Modern Physics} 2550143}
  [\href{https://arxiv.org/abs/https://doi.org/10.1142/S0219887825501439}{{\ttfamily
  https://doi.org/10.1142/S0219887825501439}}].

\bibitem{Naseer:2025ghn}
T.~Naseer, J.~Levi~Said, M.~Sharif and A.-H.~Abdel-Aty, \emph{{Thermodynamic
  properties of non-singular Hayward black hole through the lens of minimal
  gravitational decoupling}},
  \href{https://doi.org/10.1140/epjc/s10052-025-14186-x}{\emph{Eur. Phys. J. C}
  {\bfseries 85} (2025) 471}.

\bibitem{Naseer:2024tgt}
T.~Naseer, \emph{{Role of Rastall gravity in constructing new spherically
  symmetric stellar solutions}},
  \href{https://doi.org/10.1016/j.dark.2024.101663}{\emph{Phys. Dark Univ.}
  {\bfseries 46} (2024) 101663}.

\bibitem{Murtaza:2024ylz}
G.~Murtaza, A.~Ditta, T.~Naseer, G.~Mustafa, S.K.~Maurya, A.~Ghaffar et~al.,
  \emph{{On the evaluation of accretion process near a quantum-improved charged
  black hole}}, \href{https://doi.org/10.1016/j.jheap.2024.10.004}{\emph{JHEAp}
  {\bfseries 44} (2024) 279}.

\bibitem{Rasheed:2024aia}
B.~Rasheed, A.~Ditta, T.~Naseer, F.~Javed and G.~Mustafa, \emph{{Analyzing the
  quantum corrected adS spherically symmetric black holes with phantom global
  monopoles for thermal properties}},
  \href{https://doi.org/10.1142/S021988782450302X}{\emph{Int. J. Geom. Meth.
  Mod. Phys.} {\bfseries 22} (2025) 2450302}.

\bibitem{Ruppeiner:1979bcp}
G.~Ruppeiner, \emph{{Thermodynamics: A Riemannian geometric model}},
  \href{https://doi.org/10.1103/PhysRevA.20.1608}{\emph{Phys. Rev. A}
  {\bfseries 20} (1979) 1608}.

\bibitem{Ruppeiner:1995zz}
G.~Ruppeiner, \emph{{Riemannian geometry in thermodynamic fluctuation theory}},
  \href{https://doi.org/10.1103/RevModPhys.67.605}{\emph{Rev. Mod. Phys.}
  {\bfseries 67} (1995) 605}.

\bibitem{Mirza:2007ev}
B.~Mirza and M.~Zamani-Nasab, \emph{{Ruppeiner Geometry of RN Black Holes: Flat
  or Curved?}},
  \href{https://doi.org/10.1088/1126-6708/2007/06/059}{\emph{JHEP} {\bfseries
  06} (2007) 059} [\href{https://arxiv.org/abs/0706.3450}{{\ttfamily
  0706.3450}}].

\bibitem{Quevedo:2007mj}
H.~Quevedo, \emph{Geometrothermodynamics of black holes}, {\emph{Gen. Rel.
  Grav.} {\bfseries 40} (2008) 971}
  [\href{https://arxiv.org/abs/arXiv:0704.3102}{{\ttfamily arXiv:0704.3102}}].

\bibitem{Hendi:2015rja}
S.H.~Hendi, S.~Panahiyan, B.~Eslam~Panah and M.~Momennia, \emph{{A new approach
  toward geometrical concept of black hole thermodynamics}},
  \href{https://doi.org/10.1140/epjc/s10052-015-3701-5}{\emph{Eur. Phys. J. C}
  {\bfseries 75} (2015) 507}
  [\href{https://arxiv.org/abs/1506.08092}{{\ttfamily 1506.08092}}].

\bibitem{Belhaj:2015uwa}
A.~Belhaj, M.~Chabab, H.~El~Moumni, K.~Masmar and M.B.~Sedra, \emph{{On
  Thermodynamics of AdS Black Holes in M-Theory}},
  \href{https://doi.org/10.1140/epjc/s10052-016-3928-9}{\emph{Eur. Phys. J. C}
  {\bfseries 76} (2016) 73} [\href{https://arxiv.org/abs/1509.02196}{{\ttfamily
  1509.02196}}].

\bibitem{Bhattacharya:2017hfj}
K.~Bhattacharya and B.R.~Majhi, \emph{{Thermogeometric description of the van
  der Waals like phase transition in AdS black holes}},
  \href{https://doi.org/10.1103/PhysRevD.95.104024}{\emph{Phys. Rev. D}
  {\bfseries 95} (2017) 104024}
  [\href{https://arxiv.org/abs/1702.07174}{{\ttfamily 1702.07174}}].

\bibitem{Wei:2019uqg}
S.-W.~Wei, Y.-X.~Liu and R.B.~Mann, \emph{{Repulsive Interactions and Universal
  Properties of Charged Anti\textendash{}de Sitter Black Hole
  Microstructures}},
  \href{https://doi.org/10.1103/PhysRevLett.123.071103}{\emph{Phys. Rev. Lett.}
  {\bfseries 123} (2019) 071103}
  [\href{https://arxiv.org/abs/1906.10840}{{\ttfamily 1906.10840}}].

\bibitem{Wei:2019yvs}
S.W.~Wei, Y.X.~Liu and R.B.~Mann, \emph{Ruppeiner geometry, phase transitions,
  and the microstructure of charged ads black holes}, {\emph{Phys. Rev. D}
  {\bfseries 100} (2019) 124033}
  [\href{https://arxiv.org/abs/1909.03887}{{\ttfamily 1909.03887}}].

\bibitem{Wei:2015iwa}
S.W.~Wei and Y.X.~Liu, \emph{Insight into the microscopic structure of an ads
  black hole from a thermodynamical phase transition},
  \href{https://doi.org/10.1103/PhysRevLett.115.111302}{\emph{Phys. Rev. Lett.}
  {\bfseries 115} (2015) 111302}
  [\href{https://arxiv.org/abs/1502.00386}{{\ttfamily 1502.00386}}].

\bibitem{AR}
X.Y.~Guo, H.F.~Li, L.C.~Zhang and R.~Zhao, \emph{Microstructure and continuous
  phase transition of a reissner-nordström-ads black hole}, {\emph{Phys. Rev.
  D} {\bfseries 100} (2019) 064036}
  [\href{https://arxiv.org/abs/1901.04703}{{\ttfamily 1901.04703}}].

\bibitem{Mann2019}
S.W.~Wei, Y.X.~Liu and R.B.~Mann, \emph{Repulsive interactions and universal
  properties of charged anti–de sitter black hole microstructures},
  {\emph{Phys. Rev. Lett.} {\bfseries 123} (2019) 071103}
  [\href{https://arxiv.org/abs/1906.10840}{{\ttfamily 1906.10840}}].

\bibitem{SAdS}
Z.M.~Xu, B.~Wu and W.L.~Yang, \emph{Ruppeiner thermodynamic geometry for the
  schwarzschild-ads black hole}, {\emph{Phys. Rev. D} {\bfseries 101} (2020)
  024018} [\href{https://arxiv.org/abs/1910.12182}{{\ttfamily 1910.12182}}].

\bibitem{Cai:1998ep}
R.G.~Cai and J.H.~Cho, \emph{Thermodynamic curvature of the btz black hole},
  \href{https://doi.org/10.1103/PhysRevD.60.067502}{\emph{Phys. Rev. D}
  {\bfseries 60} (1999) 067502}
  [\href{https://arxiv.org/abs/hep-th/9803261}{{\ttfamily hep-th/9803261}}].

\bibitem{Shen:2005nu}
J.Y.~Shen, R.G.~Cai, B.~Wang and R.K.~Su, \emph{Thermodynamic geometry and
  critical behavior of black holes},
  \href{https://doi.org/10.1142/S0217751X07034064}{\emph{Int. J. Mod. Phys. A}
  {\bfseries 22} (2007) 11}
  [\href{https://arxiv.org/abs/gr-qc/0512035}{{\ttfamily gr-qc/0512035}}].

\bibitem{Davies:1978mf}
P.C.W.~Davies, \emph{Thermodynamics of black holes},
  \href{https://doi.org/10.1098/rspa.1977.0047}{\emph{Proc. Roy. Soc. Lond. A}
  {\bfseries 353} (1977) 499}.

\bibitem{Dehyadegari}
A.~Dehyadegari, A.~Sheykhi and A.~Montakhab, \emph{Critical behaviour and
  microscopic structure of charged ads black holes via an alternative phase
  space}, {\emph{Phys. Lett. B} {\bfseries 768} (2017) 235}
  [\href{https://arxiv.org/abs/arXiv:1607.05333}{{\ttfamily
  arXiv:1607.05333}}].

\bibitem{Sheykhi}
M.K.~Zangeneh, A.~Dehyadegari, A.~Sheykhi and R.B.~Mann, \emph{Microscopic
  origin of black hole reentrant phase transitions}, {\emph{Phys. Rev. D}
  {\bfseries 97} (2018) 084054}
  [\href{https://arxiv.org/abs/arXiv:1709.04432}{{\ttfamily
  arXiv:1709.04432}}].

\bibitem{Miao3}
Y.G.~Miao and Z.M.~Xu, \emph{Interaction potential and thermo-correction to the
  equation of state for thermally stable schwarzschild ads black holes},
  {\emph{Sci. China Phys. Mech. Astron.} {\bfseries 62} (2019) 010412}
  [\href{https://arxiv.org/abs/arXiv:1804.01743}{{\ttfamily
  arXiv:1804.01743}}].

\bibitem{Chen}
Y.~Chen, H.~Li and S.J.~Zhang, \emph{Microscopic explanation for black hole
  phase transitions via ruppeiner geometry: two competing mechanisms},
  {\emph{arXiv} (2018) }
  [\href{https://arxiv.org/abs/arXiv:1812.11765}{{\ttfamily
  arXiv:1812.11765}}].

\bibitem{hermann}
R.~Hermann, \emph{Geometry, Physics, and Systems}, Marcel Dekker (1973).

\bibitem{mrugala1996}
R.~Mrugala, \emph{On a riemannian metric on contact thermodynamic spaces},
  {\emph{Rep. Math. Phys.} {\bfseries 38} (1996) }.

\bibitem{contactBH}
A.~Ghosh and C.~Bhamidipati, \emph{Contact geometry and thermodynamics of black
  holes in ads spacetimes}, {\emph{Phys. Rev. D} {\bfseries 100} (2019) 126020}
  [\href{https://arxiv.org/abs/arXiv:1909.11506}{{\ttfamily
  arXiv:1909.11506}}].

\bibitem{Quevedo:2006xk}
H.~Quevedo, \emph{Geometrothermodynamics}, {\emph{J. Math. Phys.} {\bfseries
  48} (2007) 013506}
  [\href{https://arxiv.org/abs/arXiv:physics/0604164}{{\ttfamily
  arXiv:physics/0604164}}].

\bibitem{Zhang:2015ova}
J.L.~Zhang, R.G.~Cai and H.~Yu, \emph{Phase transition and thermodynamical
  geometry of reissner-nordström-ads black holes in extended phase space},
  {\emph{Phys. Rev. D} {\bfseries 91} (2015) 044028}
  [\href{https://arxiv.org/abs/arXiv:1502.01428}{{\ttfamily
  arXiv:1502.01428}}].

\bibitem{Hendi:2015xya}
S.H.~Hendi, A.~Sheykhi, S.~Panahiyan and B.E.~Panah, \emph{Phase transition and
  thermodynamic geometry of einstein-maxwell-dilaton black holes}, {\emph{Phys.
  Rev. D} {\bfseries 92} (2015) 064028}
  [\href{https://arxiv.org/abs/arXiv:1509.08593}{{\ttfamily
  arXiv:1509.08593}}].

\bibitem{Ghosh:2023khd}
A.~Ghosh and C.~Bhamidipati, \emph{{Contact and metric structures in black hole
  chemistry}}, \href{https://doi.org/10.3389/fphy.2023.1132712}{\emph{Front. in
  Phys.} {\bfseries 11} (2023) 1132712}
  [\href{https://arxiv.org/abs/2302.04467}{{\ttfamily 2302.04467}}].

\bibitem{Bravetti:2012dn}
A.~Bravetti and F.~Nettel, \emph{{Thermodynamic curvature and ensemble
  nonequivalence}},
  \href{https://doi.org/10.1103/PhysRevD.90.044064}{\emph{Phys. Rev. D}
  {\bfseries 90} (2014) 044064}
  [\href{https://arxiv.org/abs/1208.0399}{{\ttfamily 1208.0399}}].

\bibitem{Dolan:2015xta}
B.P.~Dolan, \emph{{Intrinsic curvature of thermodynamic potentials for black
  holes with critical points}},
  \href{https://doi.org/10.1103/PhysRevD.92.044013}{\emph{Phys. Rev. D}
  {\bfseries 92} (2015) 044013}
  [\href{https://arxiv.org/abs/1504.02951}{{\ttfamily 1504.02951}}].

\bibitem{Singh:2020tkf}
A.~Singh, A.~Ghosh and C.~Bhamidipati, \emph{{Thermodynamic curvature of AdS
  black holes with dark energy}},
  \href{https://doi.org/10.3389/fphy.2021.631471}{\emph{Front. in Phys.}
  {\bfseries 9} (2021) 65} [\href{https://arxiv.org/abs/2002.08787}{{\ttfamily
  2002.08787}}].

\bibitem{Singh:2023hit}
A.~Singh, P.~Mukherjee and C.~Bhamidipati, \emph{{Thermodynamic curvature of
  charged black holes with AdS2 horizons}},
  \href{https://doi.org/10.1103/PhysRevD.108.106011}{\emph{Phys. Rev. D}
  {\bfseries 108} (2023) 106011}
  [\href{https://arxiv.org/abs/2307.11641}{{\ttfamily 2307.11641}}].

\bibitem{Singh:2023ufh}
A.~Singh, \emph{{Thermodynamic geometry of dyonic black holes in AdS in
  extended phase space}},
  \href{https://doi.org/10.1142/S0217732323501730}{\emph{Mod. Phys. Lett. A}
  {\bfseries 38} (2023) 2350173}.

\bibitem{Singh:2025ueu}
A.~Singh and S.~Mahish, \emph{{Criticality of charged AdS black holes with
  string clouds in boundary conformal field theory}},
  \href{https://arxiv.org/abs/2504.20486}{{\ttfamily 2504.20486}}.

\bibitem{a19}
S.-W.~Wei, Y.-X.~Liu and R.B.~Mann, \emph{{Black hole solutions as topological
  thermodynamic defects}},
  \href{https://doi.org/10.1103/PhysRevLett.129.191101}{\emph{Phys. Rev. Lett.}
  {\bfseries 129} (2022) 191101}.

\bibitem{a20}
S.-W.~Wei and Y.-X.~Liu, \emph{{Topology of black hole thermodynamics}},
  \href{https://doi.org/10.1103/PhysRevD.105.104003}{\emph{Phys. Rev. D}
  {\bfseries 105} (2022) 104003}.

\bibitem{22a}
J.~Sadeghi et~al., \emph{{Bardeen black hole thermodynamics from topological
  perspective}}, \href{https://doi.org/10.1016/j.aop.2023.169391}{\emph{Annals
  Phys.} {\bfseries 455} (2023) 169391}.

\bibitem{23a}
D.~Wu, \emph{{Topological classes of thermodynamics of the four-dimensional
  static accelerating black holes}},
  \href{https://doi.org/10.1103/PhysRevD.108.084041}{\emph{Phys. Rev. D}
  {\bfseries 108} (2023) 084041}.

\bibitem{25a}
J.~Sadeghi et~al., \emph{{Bulk-boundary and RPS Thermodynamics from Topology
  perspective}}, {\emph{Chin. Phys. C} (2024) }.

\bibitem{27a}
Y.~Sekhmani et~al., \emph{{Thermodynamic topology of Black Holes in
  F(R)-Euler-Heisenberg gravity's Rainbow}}, {\emph{Arxiv} (2024) }
  [\href{https://arxiv.org/abs/2409.04997}{{\ttfamily 2409.04997}}].

\bibitem{31a}
J.~Sadeghi et~al., \emph{Topology of hayward-ads black hole thermodynamics},
  {\emph{Physica Scripta} {\bfseries 99} (2024) 025003}.

\bibitem{33a}
J.~Sadeghi et~al., \emph{Thermodynamic topology and photon spheres in the
  hyperscaling violating black holes}, {\emph{Astroparticle Physics} {\bfseries
  156} (2024) 102920}.

\bibitem{34a}
B.~Hazarika and P.~Phukon, \emph{Thermodynamic topology of black holes in f (r)
  gravity}, {\emph{Progress of Theoretical and Experimental Physics} (2024)
  043E01}.

\bibitem{38'}
S.N.~Gashti, I.~Sakalli and B.~Pourhassan, \emph{Thermodynamic topology, photon
  spheres, and evidence for weak gravity conjecture in charged black holes with
  perfect fluid within rastall theory}, {\emph{arXiv preprint arXiv:2410.14492}
  (2024) }.

\bibitem{38a}
J.~Sadeghi and M.A.S.~Afshar, \emph{The role of topological photon spheres in
  constraining the parameters of black holes}, {\emph{Astroparticle Physics}
  (2024) 102994}.

\bibitem{38b}
M.A.S.~Afshar and J.~Sadeghi, \emph{Effective potential and topological photon
  spheres: a novel approach to black hole parameter classification},
  {\emph{arXiv preprint arXiv:2405.18798} (2024) }.

\bibitem{38c}
M.A.S.~Afshar and J.~Sadeghi, \emph{Mutual influence of photon sphere and
  non-commutative parameter in various non-commutative black holes: Part
  i-towards evidence for wgc}, {\emph{arXiv preprint arXiv:2411.09557} (2024)
  }.

\bibitem{43a}
J.~Sadeghi et~al., \emph{Bulk-boundary and rps thermodynamicsfrom topology
  perspective}, {\emph{Chinese Physics C} (2024) }.

\bibitem{44a}
J.~Sadeghi et~al., \emph{Thermodynamic topology of quantum corrected
  ads-reissner-nordstrom black holes in kiselev spacetime}, {\emph{Chinese
  Physics C} (2024) }.

\bibitem{44e}
S.N.~Gashti et~al., \emph{Thermodynamic topology of kiselev-ads black holes
  within f (r, t) gravity}, {\emph{Chinese Physics C} {\bfseries 49} (2025)
  035110}.

\bibitem{44g}
S.N.~Gashti et~al., \emph{Thermodynamic topology and photon spheres of dirty
  black holes within non-extensive entropy}, {\emph{Physics of the Dark
  Universe} (2025) 101833}.

\bibitem{44h}
M.A.S.~Afshar et~al., \emph{Topological insights into black hole
  thermodynamics: Non-extensive entropy in cft framework}, {\emph{arXiv
  preprint arXiv:2501.00955} (2025) }.

\bibitem{44i}
S.N.~Gashti, B.~Pourhassan and I.~Sakallı, \emph{Thermodynamic topology and
  phase space analysis of ads black holes through non-extensive entropy
  perspectives}, {\emph{The European Physical Journal C} {\bfseries 85} (2025)
  305}.

\bibitem{44j}
S.N.~Gashti, \emph{Topology of holographic thermodynamics within non-extensive
  entropy}, {\emph{Journal of Holography Applications in Physics} {\bfseries 4}
  (2024) 59}.

\bibitem{44k}
M.R.~Alipour et~al., \emph{Topological classification and black hole
  thermodynamics}, {\emph{Physics of the Dark Universe} {\bfseries 42} (2023)
  101361}.

\bibitem{44l}
A.~Anand and S.N.~Gashti, \emph{Universality relation and thermodynamic
  topology with three-parameter entropy model}, {\emph{Physics of the Dark
  Universe} (2025) 101916}.

\bibitem{44m}
S.N.~Gashti and B.~Pourhassan, \emph{Non-extensive entropy and holographic
  thermodynamics: Topological insights}, {\emph{European Physical Journal C}
  {\bfseries 85} (2025) 435}.

\bibitem{Sekhmani:2024vsu}
Y.~Sekhmani, S.~Noori~Gashti, M.A.S.~Afshar, M.R.~Alipour, J.~Sadeghi,
  B.~Pourhassan et~al., \emph{{Thermodynamic topology of Black Holes in
  $F(R)$-Euler-Heisenberg gravity's Rainbow}},
  \href{https://arxiv.org/abs/2409.04997}{{\ttfamily 2409.04997}}.

\end{thebibliography}\endgroup

\end{document}